\documentclass[english]{article}
\usepackage[latin9]{inputenc}
\usepackage{array}
\usepackage{float}
\usepackage{booktabs}
\usepackage{multirow}
\usepackage{amsmath}
\usepackage{amsthm}
\usepackage{amssymb}
\usepackage{graphicx}

\makeatletter

\providecommand{\tabularnewline}{\\}

\numberwithin{equation}{section}
\numberwithin{figure}{section}
\theoremstyle{plain}
\newtheorem{thm}{\protect\theoremname}
  \theoremstyle{plain}
  \newtheorem{prop}[thm]{\protect\propositionname}
  \theoremstyle{plain}
  \newtheorem{lem}[thm]{\protect\lemmaname}

\usepackage{a4wide}
\usepackage{babel}
\usepackage{fullpage}
\usepackage{authblk}
\usepackage{hyperref}
\usepackage{placeins}
\usepackage{hyphenat} 
\usepackage{fancyvrb}
\usepackage[hypcap]{caption} 

\title{Machine Learning for Pricing American Options in High-Dimensional Markovian and non-Markovian models}
\author{ \textsc{Ludovic Gouden\`ege}
\footnote{This work was supported by a public grant as part of the Investissement d'avenir project, reference ANR-11-LABX-0056-LMH, LabEx LMH.} 
\thanks{F\'ederation de Math\'ematiques de CentraleSup\'elec - CNRS FR3487, France -\texttt{ ludovic.goudenege@math.cnrs.fr}}

\and \textsc{Andrea Molent}\thanks{Dipartimento di Scienze Economiche e Statistiche, Universit\`a degli Studi di Udine, Italy - \texttt{andrea.molent@uniud.it}} 
\and \textsc{Antonino Zanette}\thanks{Dipartimento di Scienze Economiche e Statistiche, Universit\`a degli Studi di Udine, Italy - \texttt{antonino.zanette@uniud.it}}}
\date{}

\def\R{{\mathbb R}}
\def\cl#1{{\cal #1}}

\makeatother

\usepackage{babel}
  \providecommand{\lemmaname}{Lemma}
  \providecommand{\propositionname}{Proposition}
\providecommand{\theoremname}{Theorem}

\begin{document}
\maketitle

\begin{flushleft}
\rule{1\columnwidth}{1pt}
\par\end{flushleft}

\begin{flushleft}
\textbf{\large{}Abstract}
\par\end{flushleft}{\large \par}

In this paper we propose two efficient techniques which allow one
to compute the price of American basket options. In particular, we
consider a basket of assets that follow a multi-dimensional Black-Scholes
dynamics. The proposed techniques, called GPR Tree (GRP-Tree) and
GPR Exact Integration (GPR-EI), are both based on Machine Learning,
exploited together with binomial trees or with a closed formula for
integration. Moreover, these two methods solve the backward dynamic
programming problem considering a Bermudan approximation of the American
option. On the exercise dates, the value of the option is first computed
as the maximum between the exercise value and the continuation value
and then approximated by means of Gaussian Process Regression. The
two methods mainly differ in the approach used to compute the continuation
value: a single step of binomial tree or integration according to
the probability density of the process. Numerical results show that
these two methods are accurate and reliable in handling American options
on very large baskets of assets. Moreover we also consider the rough
Bergomi model, which provides stochastic volatility with memory. Despite
this model is only bidimensional, the whole history of the process
impacts on the price, and handling all this information is not obvious
at all. To this aim, we present how to adapt the GPR-Tree and GPR-EI
methods and we focus on pricing American options in this non-Markovian
framework.

\vspace{2mm}

\noindent \emph{\large{}Keywords}: Machine Learning, American Options,
Multi-dimensional Black-Scholes Model, Rough Bergomi Model, Binomial
Tree Method, Exact Integration.

\noindent\rule{1\columnwidth}{1pt}

\section{Introduction}

Pricing American options is clearly a crucial question of finance
but also a challenging one since computing the optimal exercise strategy
is not an evident task. This issue is even more exacting when the
underling of the option is a multi-dimensional process, such as a
baskets of $d$ assets, since in this case the direct application
of standard numerical schemes, such as finite difference or tree methods,
is not possible because of the exponential growth of the calculation
time and the required working memory.

Common approaches in this field can be divided in four groups: techniques
which rely on recombinant trees to discretize the underlyings (see
\cite{bally2003first}, \cite{broadie1997pricing} and \cite{jain2012pricing}),
techniques which employ regression on a truncated basis of $L^{2}$
in order to compute the conditional expectations (see \cite{longstaff2001valuing}
and \cite{tsitsiklis1999optimal}), techniques which exploit Malliavin
calculus to obtain representation formulas for the conditional expectation
(see \cite{abbas2012american}, \cite{bally2005pricing}, \cite{bouchard2004discrete},
and \cite{lions2001calcul}) and techniques which make use of duality-based
approaches for Bermudan option pricing (see \cite{haugh2004pricing},
\cite{lelong2018dual} and \cite{rogers2002monte}).

Recently, Machine Learning algorithms (Rasmussen and Williams \cite{williams2006gaussian})
and Deep Learning techniques (Nielsen \cite{nielsen2015neural}) have
found great application in this sector of option pricing. 

Neural networks are used by Kohler et al. \cite{kohler2010pricing}
to price American options based on several underlyings. Deep Learning
techniques are nowadays widely used in solving large differential
equations, which is intimately related to option pricing. In particular,
Han et al. \cite{han2018} introduce a Deep Learning-based approach
that can handle general high-dimensional parabolic PDEs. E et al.
\cite{weinan2017} propose an algorithm for solving parabolic partial
differential equations and backward stochastic differential equations
in high dimension. Beck et al. \cite{beck2017} introduce a method
for solving high-dimensional fully nonlinear second-order PDEs. As
far as American options in high dimension are concerned, Becker et
al. \cite{becker2018} develop a Deep Learning method for optimal
stopping problems which directly learns the optimal stopping rule
from Monte Carlo samples.

Also Machine Learning techniques have made their contribution. For
example, Dixon and Cr{\'e}pey present a multi-Gaussian process regression
for estimating portfolio risk, and in particular the associated CVA.
De Spiegeleer et al. \cite{de2018machine} propose to apply Gaussian
Process Regression (GPR) to predict the price of the derivatives from
a training set made of observed prices for particular combinations
of model parameters. Ludkovski \cite{Ludkovski2018} proposes to use
GPR meta-models for fitting the continuation values of Bermudan options.
Similarly, Goudenège et al. \cite{goudenege2019machine} propose the
GPR-MC, which is a backward induction algorithm that employs Monte
Carlo simulations and GPR to compute the price of American options
in very high dimension (up to 100). In the insurance context, Gan
\cite{gan2013} studies the pricing of a large portfolio of Variable
Annuities in the Black-Scholes model by using clustering and GPR.
Moreover, Gan and Lin \cite{gan2015} propose a novel approach that
combines clustering technique and GPR to efficiently evaluate policies
considering nested simulations.

In this paper we present two numerical techniques which upgrade the
GPR-MC approach by replacing the Monte Carlo based computation of
the continuation value respectively with a tree step and with an exact
integration step. In particular, the algorithms we propose proceed
backward over time and compute the price function only on a set of
predetermined points. At each time step, a binomial tree step or a
closed formula for integration are used together with GPR to approximate
the continuation value at these points. The option price is then obtained
as the maximum between the continuation value and the intrinsic value
of the option and the algorithms proceed backward. For the sake of
simplicity, we name these new approaches Gaussian Process Regression
- Tree (GPR-Tree) and Gaussian Process Regression - Exact Integration
(GPR-EI). We observe that the use of the GPR method to extrapolate
the option value is particularly efficient in terms of computing time
with respect to other techniques such as Neural Networks, especially
because a small dataset is considered here. Moreover, Le Gratiet et
Garnier \cite{gratiet2012regularity} developed recent convergence
results about GPR, extending the outcomes of Rasmussen and Williams
\cite{williams2006gaussian}, and founding the convergence rate when
different kernels are employed. 

In order to demonstrate the wide applicability of the GPR methods,
we also consider the rough Bergomi model, which is a non-Markovian
model with stochastic volatility. Such a model, introduced by Bayer
et al. \cite{bayer2016pricing} stood out for explaining implied volatility
smiles and other phenomena in the pricing of European options. The
non-Markovian property of the model makes it difficult to implement
a methodologically correct approach to address the valuation of American
options. The literature in this framework is really poor. Horvat et
al. \cite{horvath2017functional} propose an approach based on Donsker\textquoteright s
approximation for fractional Brownian motion and on a tree with exponential
complexity. More recently, Bayer et al. \cite{bayer2018pricing} introduce
a method based on Monte Carlo simulation and exercise Rate Optimization. 

Numerical results show that both the GPR-Tree and the GPR-EI methods
are accurate and reliable in the multi-dimensional Black-Scholes model.
Moreover the computational times with respect to the GPR-MC method
are improved. The GPR-Tree and the GPR-EI methods prove its accuracy
also when applied to the rough Bergomi model.

The reminder of the paper is organized as follows. Section 2 presents
American options in the multi-dimensional Black-Scholes model. Section
3 and Section 4 introduce the GPR-Tree and the GPR-EI methods for
the multi-dimensional Black-Scholes model respectively. Section 5
presents the American options in the rough Bergomi model. Section
6 and Section 7 introduce the GPR-Tree and the GPR-EI methods for
the rough Bergomi model. Section 8 reports some numerical results.
Section 9 draws some conclusions. 

\section{American options in the multi-dimensional Black-Scholes model}

An American option  with maturity $T$ is a derivative instrument
whose holder can exercise the intrinsic optionality at any moment
before maturity. Let $\mathbf{S}=(\mathbf{S}_{t})_{t\in[0,T]}$ denote
the $d$-dimensional underlying  process, which is supposed to randomly
evolve according to the multi-dimensional Black-Scholes model: under
the risk neutral probability, such a model is given by the following
equation
\begin{equation}
dS_{t}^{i}=r\,S_{t}^{i}\,dt+\sigma_{i}\,S_{t}^{i}\,dW_{t}^{i},\quad\ i=1,\ldots,d,\label{sde}
\end{equation}
 with $\mathbf{S}_{0}=\left(s_{0}^{1},\dots,s_{0}^{d}\right)\in\mathbb{R}_{+}^{d}$
the spot price, $r$ the (constant) interest rate,  $\mathbf{\sigma}=(\sigma_{1},\dots,\sigma_{d})$
 the vector of volatilities, $\mathbf{W}$ a $d$-dimensional correlated
Brownian motion and $\rho_{ij}$ the instantaneous correlation coefficient
between $W_{t}^{i}$ and $W_{t}^{j}.$ Moreover, let $\Psi(\mathbf{S}_{T})$
denote the cash-flow associated with the option at maturity $T$.
Thus, the price at time $t$ of an American option having maturity
$T$ and payoff function $\Psi\,:\,\R_{+}^{d}\to\R$ is then 
\begin{equation}
v(t,\mathbf{x})=\sup_{\tau\in\mathcal{T}_{t,T}}\mathbb{E}_{t,\mathbf{x}}\left[e^{-r(\tau-t)}\Psi(\mathbf{S}_{\tau})\right],\label{price}
\end{equation}
 	 where $\cl T_{t,T}$ stands for the set of all the stopping times
taking values on $[t,T]$ and $\mathbb{E}_{t,\mathbf{x}}\left[\cdot\right]$
represents the expectation given all the information at time $t$
and in particular assuming $\mathbf{S}_{t}=\mathbf{x}$.

For simulation purposes, the $d-$dimensional Black-Scholes model
can be written alternatively using the Cholesky decomposition.  Specifically,
for $i=1,\dots,d$ we can write 
\begin{equation}
dS_{t}^{i}=S_{t}^{i}(rdt+\sigma_{i}\Sigma_{i}d\mathbf{B}_{t}),\label{sde_cho}
\end{equation}
 where $\mathbf{B}$ is a $d$-dimensional uncorrelated Brownian motion
and $\Sigma_{i}$ is the $i$-th row of the matrix $\Sigma$ defined
as a square root of the correlation matrix $\Gamma$, given by 
\begin{equation}
\Gamma=\begin{pmatrix}1 & \rho_{12} & \hdots & \rho_{1d}\\
\rho_{21} & 1 & \ddots & \vdots\\
\vdots & \ddots & \ddots & \vdots\\
\rho_{d1} & \hdots & \hdots & 1
\end{pmatrix}
\end{equation}

\section{The GPR-Tree method in the multi-dimensional Black-Scholes model}

The GPR-Tree method is similar to the GPR-MC method but the diffusion
of the underlyings is performed through a step of a binomial tree.
In particular, the algorithm proceeds backward over time, approximating
the price of the American option with the price of a Bermudan option
on the same basket. At each time step, the price function is evaluated
only on a set of predetermined points, through a binomial tree step
together with GPR to approximate the continuation value. Finally,
the optionality is exploited by computing the option value as the
maximum between the continuation value and the exercise value. 

Let $N$ denote the number of time steps, $\Delta t=T/N$ be the time
increment and $t_{n}=n\,\Delta t$ represent the discrete exercise
dates for $n=0,1,\ldots,N$. At any exercise date $t_{n}$, the value
of the option is determined by the vector of the underlying prices
$\mathbf{S}_{t_{n}}$ as follows:
\begin{equation}
v\left(t_{n},\mathbf{S}_{t_{n}}\right)=\max\left(\Psi\left(\mathbf{S}_{t_{n}}\right),C\left(t_{n},\mathbf{S}_{t_{n}}\right)\right),\label{eq:update}
\end{equation}
where $C$ denotes the continuation value of the option and it is
given by the following relation: 
\begin{equation}
C\left(t_{n},\mathbf{S}_{t_{n}}\right)=\mathbb{E}_{t_{n},\mathbf{S}_{t_{n}}}\left[e^{-r\Delta t}v\left(t_{n+1},\mathbf{S}_{t_{n+1}}\right)\right].\label{eq:CV}
\end{equation}

We observe that, if the function $v\left(t_{n+1},\cdot\right)$ is
known, then it is possible to compute $v\left(t_{n},\cdot\right)$
by approximating the expectation in (\ref{eq:CV}). In order to obtain
such an approximation, we consider a set $X$ of $P$ points whose
elements represent certain possible values for the underlyings $\mathbf{S}$:
\begin{equation}
X=\left\{ \mathbf{x}^{p}=\left(x_{1}^{p},\dots,x_{d}^{p}\right),p=1,\dots,P\right\} \subset\mathbb{R}_{+}^{d},\label{eq:X}
\end{equation}
where $\mathbb{R}_{+}^{d}=\left]0,+\infty\right[^{d}$. Such a set
is determined as done by Goudenège et al. \cite{goudenege2019machine},
that is the elements of $X$ are obtained through a quasi-random simulation
of $\mathbf{S}_{T}$ based on the Halton sequence (see \cite{goudenege2019machine}
for more details). 

The GPR-Tree method assesses $v\left(t_{n},\mathbf{x}^{p}\right)$
for each $\mathbf{x}^{p}\in X$ through one step of the binomial tree
proposed by Ekval \cite{ekvall1996lattice}. In particular, for each
$\mathbf{x}^{p}\in X$, we consider a set $\tilde{X}^{p}$ of $2^{d}$
possible values for $\mathbf{S}_{t_{n+1}}$ 
\begin{equation}
\tilde{X}^{p}=\left\{ \mathbf{\tilde{x}}^{p,k}=\left(\tilde{x}_{1}^{p,k},\dots,\tilde{x}_{d}^{p,k}\right),k=1,\dots,2^{d}\right\} \subset\mathbb{R}_{+}^{d}
\end{equation}
 which are computed as follows: 
\begin{equation}
\mathbf{\tilde{x}}_{i}^{p,k}=\mathbf{x}_{i}^{p}\exp\left(\left(r-\frac{\sigma_{i}^{2}}{2}\right)\Delta t+\sigma_{i}\sqrt{\Delta t}\Sigma_{i}\mathbf{G}_{k}\right),\ k=1,\dots,2^{d}
\end{equation}
being $\mathbf{G}_{k}$ the $k$-th point of the space $\left\{ -1,+1\right\} ^{d}$.
In particular, if $\mathbf{Y}_{k}\in\left\{ 0,1\right\} ^{d}$ is
the vector whose components are the digits of the binary representation
of $2^{d}-1$, then $\mathbf{G}_{k}=2\mathbf{Y}_{k}-1$. It is worth
noticing that, as pointed out in \cite{ekvall1996lattice}, the elements
of $\tilde{X}^{p}$ are equally likely and this simplifies the evaluation
of the expected value to the computation of the arithmetic mean of
the future values. Using the tree step, the price function may be
approximated by
\begin{equation}
v_{n}^{Tree}\left(\mathbf{x}^{p}\right)=\max\left(\Psi\left(\mathbf{x}^{p}\right),\frac{e^{-r\Delta t}}{2^{d}}\sum_{k=1}^{2^{d}}v\left(t_{n+1},\mathbf{\tilde{x}}^{p,k}\right)\right).\label{eq:update2}
\end{equation}
The computation in (\ref{eq:update2}) can be performed only if the
quantities $v\left(t_{n+1},\mathbf{\tilde{x}}^{p,k}\right)$ are known
for all the future points $\mathbf{\tilde{x}}^{p,k}$. If we proceed
backward, the function $v\left(t,\cdot\right)$ is known at maturity
since it is given by the payoff function $\Psi\left(\cdot\right)$
and so (\ref{eq:update2}) can be computed at $t_{N-1}$ and for all
the points of $X$. In order to compute $v\left(t_{N-2},\mathbf{x}^{p}\right)$
for all $\mathbf{x}^{p}\in X$, and thus going on up to $t=0$, we
have to evaluate the function $v\left(t_{N-1},\cdot\right)$ for all
the points in $\tilde{X}=\bigcup_{p=1}^{P}\tilde{X}^{p}$, but we
only know $v_{N-1}^{Tree}\left(\cdot\right)$ at $X$. To overcome
this issue, we employ the GPR method to approximate the function $v_{N-1}^{Tree}\left(\cdot\right)$
at any point of $\mathbb{R}^{d}$ and in particular at the elements
of $\tilde{X}$. Specifically, let $v_{N-2}^{GPR}\left(\cdot\right)$
denote the GPR prediction of $v_{N-1}^{Tree}\left(\cdot\right)$,
obtained by considering the predictor set $X$ and the response $\mathbf{y}\in\mathbb{R}^{P}$
given by 
\begin{equation}
y^{p}=v_{N-1}^{Tree}\left(\mathbf{x}^{p}\right),\ p\in\left\{ 1,\dots,P\right\} .
\end{equation}
The GPR-Tree approximation $v_{N-2}^{GPR-Tree}\left(\cdot\right)$
of the value function $v\left(t_{N-2},\cdot\right)$ at time $t_{N-2}$
can be computed as follows:

\begin{equation}
v_{N-2}^{GPR-Tree}\left(\mathbf{x}^{p}\right)=\max\left(\Psi\left(\mathbf{x}^{p}\right),\frac{e^{-r\Delta t}}{2^{d}}\sum_{k=1}^{2^{d}}v_{N-1}^{GPR}\left(\mathbf{\tilde{x}}^{p,k}\right)\right),\ p\in\left\{ 1,\dots,P\right\} .
\end{equation}
Following the same steps, the dynamic programming problem can be solved.
Specifically, let $n\in\left\{ 0,\dots,N-3\right\} $ and let $v_{n+1}^{GPR}\left(\cdot\right)$
denote the GPR prediction of $v_{n+1}^{GPR-Tree}\left(\cdot\right)$
obtained from predictor set $X$ and the response $\mathbf{y}\in\mathbb{R}^{P}$
given by 
\begin{equation}
y^{p}=v_{n+1}^{GPR-Tree}\left(\mathbf{x}^{p}\right).
\end{equation}
Then, the function $v_{n}^{GPR-Tree}$ can be obtained as

\begin{equation}
v_{n}^{GPR-Tree}\left(\mathbf{x}^{p}\right)=\max\left(\Psi\left(\mathbf{x}^{p}\right),\frac{e^{-r\Delta t}}{2^{d}}\sum_{k=1}^{2^{d}}v_{n+1}^{GPR}\left(\mathbf{\tilde{x}}^{p,k}\right)\right).
\end{equation}

\section{The GPR-EI method in the multi-dimensional Black-Scholes model}

The GPR-EI method differs from both the GPR-MC and GPR-Tree methods
for two reasons. First of all, the predictors employed in the GPR
step are related to the logarithms of the predictors used in the GPR-Tree
method. Secondly, the continuation value at these points is computed
through a closed formula which comes from an exact integration. 

Let $X=\left\{ \mathbf{x}^{p},p=1,\dots,P\right\} \subset\mathbb{R}_{+}^{d}$
be the same set as in (\ref{eq:X}) and define $\log\left(\mathbf{x}^{p}\right)$
as the vector obtained by applying the natural logarithm to all the
components of $\mathbf{x}^{p}$, that is $\log\left(\mathbf{x}^{p}\right)=\left(\log\left(x_{1}^{p}\right),\dots,\log\left(x_{d}^{p}\right)\right)^{\top}$.
Moreover, let us define the set 
\begin{equation}
Z=\left\{ \mathbf{z}^{p}=\log\left(\mathbf{x}^{p}\right)-\left(r-\frac{1}{2}\boldsymbol{\sigma}^{2}\right)T,p=1,\dots,P\right\} .\label{eq:Z}
\end{equation}
In this case, we do not work directly with the function $v$, but
we rather consider the function $u:\left[0,T\right]\times Z\rightarrow\mathbb{R}$
defined as 
\begin{equation}
u\left(t,\mathbf{z}\right):=v\left(t,\exp\left(\mathbf{z}+\left(r-\frac{1}{2}\boldsymbol{\sigma}^{2}\right)t\right)\right).\label{eq:u_def}
\end{equation}
In a nutshell, the main idea is to approximate the function $u$ at
$t_{N},t_{N-1},\dots,t_{1}$ by using the GPR method on the fixed
grid $Z$. In particular, we employ the Squared Exponential Kernel
$k_{SE}:\mathbb{R}^{d}\times\mathbb{R}^{d}\rightarrow\mathbb{R}$,
which is given by
\begin{equation}
k_{SE}\left(\mathbf{a},\mathbf{b}\right)=\sigma_{f}^{2}\exp\left(-\frac{\left(\mathbf{a}-\mathbf{b}\right)^{\top}I_{d}\left(\mathbf{a}-\mathbf{b}\right)}{2\sigma_{l}^{2}}\right),\ \mathbf{a},\mathbf{b}\in\mathbb{R}^{d},\label{eq:A11-1}
\end{equation}
where $I_{d}$ the $d\times d$ identity matrix, $\sigma_{l}\in\mathbb{R}$
is the characteristic length scale and $\sigma_{f}\in\mathbb{R}$
is the signal standard deviation. These two parameters are obtained
by means of a maximum likelihood estimation. The GPR approach allows
one to approximate the function $u\left(t_{n},\cdot\right)$ at time
$t_{n}$ by

\begin{equation}
u_{n}^{GPR}\left(\mathbf{z}\right)=\sum_{q=1}^{P}k_{SE}\left(\mathbf{z}^{q},\mathbf{z}\right)\mathbf{\omega}_{q},
\end{equation}
where $\omega_{1},\dots,\omega_{P}$ are weights that are computed
by solving a linear system. The continuation value can be computed
by integrating the function $u^{GPR}$ against a $d$-dimensional
probability density. This calculation can be done easily by means
of a closed formula. 

Specifically, the GPR-EI method relies on the following Proposition.
\begin{prop}
\label{prop:0}Let $n\in\left\{ 0,\dots,N-1\right\} $ and suppose
the function $u\left(t_{n+1},\cdot\right)$ at time $t_{n+1}$ to
be known at $Z$. The GPR-EI approximation of the option value $u\left(t_{n},\cdot\right)$
at time $t_{n}$ at $\mathbf{z}^{p}$ is given by

\begin{equation}
u_{n}^{GPR-EI}\left(\mathbf{z}^{p}\right)=\max\left(\Psi\left(\exp\left(\mathbf{z}^{p}+\left(r-\frac{1}{2}\boldsymbol{\sigma}^{2}\right)t_{n}\right)\right),e^{-r\Delta t}\sum_{q=1}^{P}\omega_{q}\sigma_{f}^{2}\sigma_{l}^{d}\frac{e^{-\frac{1}{2}\left(\mathbf{z}^{q}-\mathbf{z}^{p}\right)^{\top}\left(\Pi+\sigma_{l}^{2}I_{d}\right)^{-1}\left(\mathbf{z}^{q}-\mathbf{z}^{p}\right)}}{\sqrt{\det\left(\Pi+\sigma_{l}^{2}I_{d}\right)}}\right)\label{eq:GPR-EI_0}
\end{equation}
 $\sigma_{f}$, $\sigma_{l}$, and $\omega_{1},\dots,\omega_{P}$
are certain constants determined by the GPR approximation of the function
$\mathbf{z}\mapsto u\left(t_{n+1},\mathbf{z}\right)$ for $k=1,\dots,P$,
considering $Z$ as the predictor set, and $\Pi=\left(\Pi_{i,j}\right)$
is the $d\times d$ covariance matrix of the log-increments defined
by $\Pi_{i,j}=\rho_{i,j}\sigma_{i}\sigma_{j}\Delta t$.
\end{prop}

The proof of Proposition \ref{prop:0} is reported in the Appendix
\ref{ApA0}. Equation (\ref{eq:GPR-EI_0}) allows one to compute the
option price at time $t=0$ by proceeding backward. In fact, the function
$u\left(t_{N},\cdot\right)$ is known at time $t_{N}=T$ throught
(\ref{eq:u_def}) since the price function $v\left(t_{N},\cdot\right)$
is equal to the payoff function $\Psi\left(\cdot\right)$. Moreover,
if an approximation of $u\left(t_{n+1},\cdot\right)$ is available,
then one can approximate $u\left(t_{n},\cdot\right)$ at $Z$ by means
of relation (\ref{eq:GPR-EI_0}). Finally, the option price at time
$t=0$ is approximated by $u_{0}^{GPR-EI}\left(\log\left(\mathbf{\mathbf{S}_{0}}\right)\right)$.

\section{American options in the rough Bergomi model}

The rough Bergomi model, introduced by Bayer et al. \cite{bayer2016pricing},
shapes the underlying process $S_{t}$ and its volatility $V_{t}$
through the following relations:

\begin{align}
dS_{t} & =rS_{t}dt+\sqrt{V_{t}}S_{t}dW_{t}^{1}\\
V_{t} & =\xi_{0}\left(t\right)\exp\left(\eta\widetilde{W}_{t}^{H}-\frac{1}{2}\eta^{2}t^{2H}\right),
\end{align}
with $r$ the (constant) interest rate, $\eta$ a positive parameter
and $H\in\left(0,1\right)$ the Hurst parameter. The deterministic
function $\xi_{0}\left(t\right)$ represents the forward variance
curve and following Bayer et al. \cite{bayer2016pricing} we consider
it as constant. The process $W_{t}^{1}$ is a Brownian motion, whereas
$\widetilde{W}_{t}^{H}$ is a Riemann-Liouville fractional Brownian
motion that can be expressed as a stochastic integral:
\begin{equation}
\widetilde{W}_{t}^{H}=\sqrt{2H}\int_{0}^{t}\left(t-s\right)^{H-\frac{1}{2}}dW_{t}^{2},
\end{equation}
with $W_{t}^{2}$ a Brownian motion and $\rho$ the instantaneous
correlation coefficient between $W_{t}^{1}$ and ~$W_{t}^{2}$.

The rough Bergomi model stood out for its ability to explain implied
volatility and other phenomena related to European options. Moreover,
it is particularly interesting from a computational point of view
as it is a non-Markovian model and therefore it is not possible to
apply standard techniques for American options.

In this framework, the price at time $t$ of an American option having
maturity $T$ and payoff function $\Psi\,:\,\R_{+}\to\R$ is then
\begin{equation}
v(t,\mathcal{F}_{t})=\sup_{\tau\in\mathcal{T}_{t,T}}\mathbb{E}\left[e^{-r(\tau-t)}\Psi(S_{\tau})|\mathcal{F}_{t}\right],\label{r-price}
\end{equation}
where $\mathcal{F}_{t}$ is the natural filtration generated by the
couple $\left(W_{s}^{1},\widetilde{W}_{s}^{H}\right)$ for $s\in\left[0,t\right]$.
We point out that, as opposed to the multi-dimensional Brownian motion,
in this case, the stopping time $\tau$ does not only depend from
the actual values of $S$ and $V$ but, since these are non-Markovian
processes, it depends on the whole filtration, that is from the whole
observed history of the processes. 

\section{The GPR-Tree method in the rough Bergomi model}

The GPR-Tree method can be adapted to price American options in the
rough Bergomi model. Despite the dimension of the model is only two,
it is a non-Markovian model which obliges one to take into account
the past history when evaluating the price of an option. So, the price
of an option at a certain moment depends on all the filtration at
that moment. Clearly, evaluating an option by considering the whole
history of the process (a continuous process) is not possible. To
overcome such an issue, we simulate the process on a finite number
of dates and we consider the sub-filtration induced by these observations.
First of all, we consider a finite number $N$ of time steps that
determines the time increment $\Delta t=\frac{T}{N}$, and we employ
the scheme presented in Bayer et. al \cite{bayer2016pricing} to generate
a set of $P$ simulations of the couple $\left(S_{t},V_{t}\right)$
at $t_{n}=n\,\Delta t$ for $n=1,\ldots,N$. In particular, if we
set $\Delta W_{n}^{1}=W_{t_{n}}^{1}-W_{t_{n-1}}^{1}$, then the $2N$-dimensional
random vector $\mathbf{R}$, given by
\begin{equation}
\mathbf{R}=\left(\Delta W_{1}^{1},\widetilde{W}_{t_{1}}^{H},\dots,\Delta W_{N}^{1},\widetilde{W}_{t_{N}}^{H}\right)^{\top},\label{eq:vector_R}
\end{equation}
follows a zero-mean Gaussian distribution. Moreover, using the relations
stated in Appendix \ref{ApA}, one can calculate the covariance matrix
$\Upsilon$ of $\mathbf{R}$ and its lower triangular square root
$\Lambda$ by using the Cholesky factorization. The vector $\mathbf{R}$
can be simulated by computing $\Lambda\mathbf{G}$, where $\mathbf{G}=\left(G_{1},\dots,G_{2N}\right)^{\top}$
is a vector of independent standard Gaussian random variables. Finally,
a simulation for $\left(S_{t_{n}},V_{t_{n}}\right)_{n=0,\dots,N}$
can be obtained from $\mathbf{R}$ by considering the initial values
\begin{equation}
S_{t_{0}}=S_{0},\ V_{t_{0}}=\xi_{0},\label{eq:62}
\end{equation}
and the Euler\textendash Maruyama scheme given by
\begin{align}
S_{t_{n+1}} & =S_{t_{n}}\exp\left(\left(r-\frac{1}{2}V_{t_{n}}\right)\Delta t+\sqrt{V_{t_{n}}}\Delta W_{n+1}^{1}\right),\label{eq:63}\\
V_{t_{n+1}} & =\xi_{0}\exp\left(-\frac{1}{2}\eta^{2}\left(t_{n+1}\right)^{2H}+\eta\widetilde{W}_{t_{n+1}}^{H}\right).\label{eq:64}
\end{align}

First of all, the GPR-Tree method simulates $P$ different samples
for the vector $\mathbf{G}$, namely $\mathbf{G}^{p}$ for $p=1,\dots,P$,
and it computes the corresponding paths $\left(S_{t_{1}}^{p},V_{t_{1}}^{p},\dots,S_{t_{N}}^{p},V_{t_{N}}^{p}\right)$
according to (\ref{eq:62}), (\ref{eq:63}) and (\ref{eq:64}). To
summarize the values assumed by $S$ and $V$, let us define the vector
\begin{equation}
\mathbf{SV}_{i:j}^{p}=\left(S_{t_{i}}^{p},V_{t_{i}}^{p},S_{t_{i+1}}^{p},V_{t_{i+1}}^{p},\dots,S_{t_{j}}^{p},V_{t_{j}}^{p}\right)^{\top}
\end{equation}
 for $i,j\in\left\{ 0,\dots,N\right\} $ and $i<j$. Moreover, we
also define 
\begin{equation}
\log\left(\mathbf{SV}_{i:j}^{p}\right)=\left(\log\left(S_{t_{i}}^{p}\right),\log\left(V_{t_{i}}^{p}\right),\log\left(S_{t_{i+1}}^{p}\right),\log\left(V_{t_{i+1}}^{p}\right),\dots,\log\left(S_{t_{j}}^{p}\right),\log\left(V_{t_{j}}^{p}\right)\right)^{\top},
\end{equation}
where $\log$ stands for the natural logarithm.

Then, the GPR-Tree method computes the option value for each of these
$P$ trajectories, proceeding backward in time and considering the
past history coded into the filtration. Since we consider only a finite
number of steps, we approximate the filtration $\mathcal{F}_{t_{n}}$
with the natural filtration $\hat{\mathcal{F}_{t_{n}}}$ generated
by the $2n$ variables $W_{t_{1}}^{1},\widetilde{W}_{t_{1}}^{H},\dots,W_{t_{n}}^{1},\widetilde{W}_{t_{n}}^{H}$.
Moreover, $\hat{\mathcal{F}}_{t_{n}}$ is equal to the filtration
generated by $S_{t_{1}},V_{t_{1}},\dots,S_{t_{n}},V_{t_{n}}$ because
there exists a deterministic bijective function that allows one to
obtain $W_{t_{1}}^{1},\widetilde{W}_{t_{1}}^{H},\dots,W_{t_{n}}^{1},\widetilde{W}_{t_{n}}^{H}$
from $S_{t_{1}},V_{t_{1}},\dots,S_{t_{n}},V_{t_{n}}$ and vice versa.
Therefore, when we calculate the option value conditioned by filtration
$\hat{\mathcal{F}}_{t_{n}}$, it is enough to conditioning with respect
to the knowledge of the variables $S_{t_{1}},V_{t_{1}},\dots,S_{t_{n}},V_{t_{n}}$.

The GPR-Tree method proceeds backward in time, using a tree method
and the GPR to calculate the option price with respect to the initially
simulated trajectories. As opposed to the multi-dimensional Black-Scholes
model, here we perform more than one single tree step, so as to reduce
the number of GPR regressions and thus increasing the computational
efficiency. In particular, we consider $N=N^{Tree}\cdot m$ with $N^{Tree}$
and $m$ natural numbers that represent how many times the tree method
is used and the number of time steps employed, respectively.

After simulating the $P$ random paths $\left\{ \mathbf{SV}_{1:N}^{p},\ p=1,\dots,P\right\} $,
we compute the tree approximation of the option value $v\left(t_{N-m},\mathbf{SV}_{1:\left(N-m\right)}^{p}\right)$
at time $t_{N-m}$ for each path as follows:
\begin{equation}
v_{N-m}^{Tree}\left(\mathbf{SV}_{1:\left(N-m\right)}^{p}\right)=\max\left(\Psi\left(S_{t_{N-m}}^{p}\right),C_{N-m}^{Tree}\left(\mathbf{SV}_{1:\left(N-m\right)}^{p}\right)\right),
\end{equation}
with $C_{N-m}^{Tree}$ stands for the the approximation of the continuation
value function at time $t_{N-m}$ obtained by means of a tree approach,
which discretizes each component of the Gaussian vector $\mathbf{G}_{\left[2\left(N-m\right)+1\right]:2N}$
that generates the process. As opposed to the multi-dimensional Black-Scholes
model, the approximation of the independent Gaussian components of
$\mathbf{G}$ through the equiprobable couple $\left\{ -1,+1\right\} $
is not suitable since the convergence to the right price is too slow.
So, we propose to use the same discrete approximation employed by
Alfonsi in \cite{alfonsi2010high}, which is stated in the following
Lemma.
\begin{lem}
\label{lem:alfonsi}The discrete variable $A$ defined by $\mathbb{P}\left(A=\sqrt{3+\sqrt{6}}\right)=\mathbb{P}\left(A=-\sqrt{3+\sqrt{6}}\right)=\frac{\sqrt{6}-2}{4\sqrt{6}}$
and $\mathbb{P}\left(A=\sqrt{3-\sqrt{6}}\right)=\mathbb{P}\left(A=-\sqrt{3-\sqrt{6}}\right)=\frac{1}{2}-\frac{\sqrt{6}-2}{4\sqrt{6}}$
fits the first seven moments of a standard Gaussian random variable.
\end{lem}

So, for each path $p$, we consider a quadrinomial tree with $m$
time steps, and we use it to compute the continuation value. In particular,
we consider the discrete time process $\left(\hat{S}_{k}^{p},\hat{V}_{k}^{p}\right)_{k\in\left\{ N-m,\dots,N\right\} }$defined
through 
\begin{equation}
\hat{S}_{N-m}^{p}=S_{t_{N-m}}^{p},\hat{V}_{N-m}^{p}=V_{t_{N-m}}^{p}
\end{equation}
\begin{align}
\hat{S}_{k+1}^{p} & =\hat{S}_{j-1}^{p}\exp\left(\left(r-\frac{1}{2}\hat{V}_{k}^{p}\right)\Delta t+\sqrt{\hat{V}_{k}^{p}}\Lambda_{2k+1}\hat{\mathbf{G}}^{p}\right),\label{eq:63-1}\\
\hat{V}_{k+1}^{p} & =\xi_{0}\exp\left(-\frac{1}{2}\eta^{2}\left(t_{k+1}\right)^{2H}+\eta\Lambda_{2k+2}\hat{\mathbf{G}}^{p}\right),\label{eq:64-1}
\end{align}
where $\Lambda_{2k+1}$ is the $2k+1$-th rows of the matrix $\Lambda$
and $\Lambda_{2k+2}$ the $2k+2$-th row. Moreover, $\hat{G}_{j}^{p}=G_{j}^{p}$
for $j=1,\dots,2\left(N-m\right)$ and the other components, that
is $\hat{G}_{j}^{p}$ for $j=2\left(N-m\right)+1,\dots,2N$, are sampled
by using the random variable $A$ of Lemma \ref{lem:alfonsi}.

An option value is assigned to each node of the tree: at maturity,
that is for $k=N,$ it is equal to the payoff $\Psi\left(\hat{S}_{N}^{p}\right)$,
and for $k=N-m,\dots,N-1$ it can be obtained as the maximum between
the exercise value and the discounted mean value at the future nodes,
weighted according to the transition probabilities determined by the
probability distribution of $A$.

This approach allows us to compute the function $v_{N-m}^{GPR-Tree}\left(\mathbf{SV}_{1:\left(N-m\right)}^{p}\right)$
for $p=1,\dots,P$. We point out that, since the quadrinomial tree
is not recombinant, the number of nodes grows exponentially with the
number of time steps $m$. Therefore, $m$ must be small. A similar
problem arises with the tree approach proposed by Horvat et al. \cite{horvath2017functional}.
In order to overcome such an issue, we apply the GPR method to approximate
the function $u_{N-m}^{GPR-Tree}\left(\log\left(\mathbf{SV}_{1:\left(N-m\right)}^{p}\right)\right)=v_{N-m}^{GPR-Tree}\left(\mathbf{SV}_{1:\left(N-m\right)}^{p}\right)$.
Specifically, consider a natural number $J$ and define $d_{n}=2\min\left(n,J+1\right)$.
We train the GPR method considering the predictor set given by
\begin{equation}
X=\left\{ \mathbf{x}^{p}=\log\left(\mathbf{SV}_{\max\left\{ 1,N-m-J\right\} :N-m}^{p}\right),p=1,\dots,P\right\} \subset\mathbb{R}^{d_{N-m}}
\end{equation}
and the response $\mathbf{y}\in\mathbb{R}^{P}$ given by 
\begin{equation}
y^{p}=v_{N-m}^{Tree}\left(\mathbf{SV}_{1:\left(N-m\right)}^{p}\right).
\end{equation}

We term $u_{N-m}^{GPR}$ the function obtained by the aforementioned
regression, which depends on\linebreak[4] $\log\left(\mathbf{SV}_{\max\left\{ 1,N-m-J\right\} :N-m}^{p}\right)$.
We stress out that if we consider $J=N-m-1$ (or greater), then the
function $u_{N-m}^{GPR}$ would consider all the observed values of
$S$ and $V$ as predictors. Anyway, numerical tests show that it
is enough to consider smaller values of $J$, which reduces the dimension
$d_{N-m}$ of the regression and thus improves the numerical efficiency.
A similar approach is taken by Bayer et al. \cite{bayer2018pricing}. 

Once we have obtained $u_{N-m}^{GPR}$, we can approximate the option
value $v\left(t_{N-2m},\mathbf{SV}_{1:\left(N-2m\right)}^{p}\right)$
at time $t_{N-2m}$ by means of the tree approach again. The only
difference in this case is that the value attributed to the terminal
nodes is not determined by the payoff function, but through the function
$u_{N-m}^{GPR}$. We term $v_{N-2m}^{GPR-Tree}$ the function obtained
after this backward tree step. If we train the GPR method considering
the predictor set given by
\begin{equation}
X=\left\{ \mathbf{x}^{p}=\log\left(\mathbf{SV}_{\max\left\{ 1,N-2m-J\right\} :N-2m}^{p}\right),p=1,\dots,P\right\} \subset\mathbb{R}^{d_{N-2m}}
\end{equation}
and the response $\mathbf{y}\in\mathbb{R}^{P}$ given by 
\begin{equation}
y^{p}=v_{N-2m}^{Tree}\left(\mathbf{SV}_{1:\left(N-2m\right)}^{p}\right),
\end{equation}
then we obtain the function $u_{N-2m}^{GPR}$, which can be employed
to repeat the tree step and the GPR step, proceeding backward up to
obtaining the initial option price by backward induction.

\section{The GPR-EI method in the rough Bergomi model}

The GPR-EI method can be adapted to price American options in the
rough Bergomi model. Just like the GPR-Tree approach, the GPR-EI method
starts by simulating $P$ different paths $\left(S_{t_{1}}^{p},V_{t_{1}}^{p},\dots,S_{t_{N}}^{p},V_{t_{N}}^{p}\right)$
for the processes $S$ and $V$, and it goes on by solving a backward
induction problem, through the use of the GPR method and a closed
formula for integration. 

As opposed to the multi-dimensional Black-Scholes model, in the rough
Bergomi case the use of the squared exponential kernel is not suitable
because it is a isotropic kernel and the predictors employed have
different nature (prices and volatilities at different times) and
thus changes in each predictor impact differently on the price. So,
we employ the Automatic Relevance Determination (ARD) Squared Exponential
Kernel, that has separate length scale for each predictor and it is
given by
\begin{equation}
k_{ASE}\left(\mathbf{a},\mathbf{b}\right)=\sigma_{f}^{2}\exp\left(-\sum_{i=1}^{d}\frac{\left(a_{i}-b_{i}\right)^{2}}{2\sigma_{i}^{2}}\right),\ \mathbf{a},\mathbf{b}\in\mathbb{R}^{d}.
\end{equation}
 with $d$ the number of the considered predictors. Specifically,
the GPR-EI method relies on the following Propositions.
\begin{prop}
\label{lem:L1}The GPR-EI approximation of the option value at time
$t_{N-1}$ at $\mathbf{SV}_{\max\left\{ 1,N-1-J\right\} :\left(N-1\right)}^{p}$
is given by:
\begin{equation}
v_{N-1}^{GPR-EI}\left(\mathbf{SV}_{\max\left\{ 1,N-1-J\right\} :\left(N-1\right)}^{p}\right)=\max\left(\Psi\left(S_{t_{N-1}}^{p}\right),\sum_{q=1}^{P}\frac{\mathbf{\omega}_{q}e^{-r\Delta t}\sigma_{f}^{2}\sigma_{l}}{\sqrt{\sigma_{N,p}^{2}+\sigma_{l}^{2}}}\exp\left(-\frac{\left(\log\left(S_{t_{N}}^{q}\right)-\mu_{N,p}\right)^{2}}{2\sigma_{N,p}^{2}+2\sigma_{l}^{2}}\right)\right)\label{eq:v_Nm1}
\end{equation}
where $\sigma_{f}$, $\sigma_{j}$, and $\omega_{1},\dots,\omega_{P}$
are certain constants determined by the GPR approximation of the function
$\log\left(S_{T}\right)\mapsto\Psi\left(S_{T}\right)$. Moreover,
\begin{equation}
\mu_{N,p}=\log\left(S_{t_{N-1}}^{p}\right)+\left(r-\frac{1}{2}\sqrt{V_{t_{N-1}}^{p}}\right)\Delta t
\end{equation}
and 
\begin{equation}
\sigma_{N,p}^{2}=V_{t_{N-1}}^{p}\Delta t.
\end{equation}
\end{prop}

The proof of Proposition \ref{lem:L1} is reported in the Appendix
\ref{ApA2}. Therefore, we can compute the value of the option at
time $t_{N-1}$ for each simulated path by using (\ref{eq:v_Nm1}).
\begin{prop}
\label{lem:L2}Let $n\in\left\{ 0,\dots,N-2\right\} $ and suppose
the option price function $v\left(t_{n+1},\cdot\right)$ at time $t_{n+1}$
to be known for all the simulated paths $\left\{ \mathbf{SV}_{1:N}^{p},p=1,\dots,P\right\} $.
Define 

\begin{equation}
\boldsymbol{\mu}_{n+1,p}=\left(\log\left(S_{t_{n}}^{p}\right)+\left(r-\frac{1}{2}V_{t_{n}}^{p}\right)\Delta t,\log\left(\xi_{0}\right)+\eta\Lambda_{2n+2}\underline{\mathbf{G}}^{p}-\frac{1}{2}\eta^{2}t_{n+1}^{2H}\right)^{\top},
\end{equation}
where $\Lambda_{2n+2}$ is the $2n+2$-th row of the matrix $\Lambda$
and $\underline{\mathbf{G}}^{p}=\left(G_{1}^{p},\dots,G_{2n}^{p},0\dots,0\right)^{\top}$,
and 
\begin{equation}
\Sigma_{n+1,p}=\left(\begin{array}{cc}
\Delta tV_{t_{n}}^{p} & \eta\sqrt{\Delta tV_{t_{n}}^{p}}\Lambda_{2n+2,2n+1}\\
\eta\sqrt{\Delta tV_{t_{n}}^{p}}\Lambda_{2n,2n+1} & \eta^{2}\left(\Lambda_{2n+2,2n+2}^{2}+\Lambda_{2n+2,2n+1}^{2}\right)
\end{array}\right),
\end{equation}
where $\Lambda_{i,j}$ stands for the element of $\Lambda$ in position
$i,j$. Moreover, consider a natural number $J\in\mathbb{N}$ and
set $d_{n+1}=2\min\left\{ n+1,J+1\right\} .$ Then, the GPR-EI approximation
of the option value at time $t_{n}$ at $\mathbf{SV}_{\max\left\{ 1,n-J\right\} :n}^{p}$
is given by
\begin{equation}
v_{n}^{GPR-EI}\left(\mathbf{SV}_{\max\left\{ 1,n-J\right\} :n}^{p}\right)=\max\left(\Psi\left(S_{t_{n}}^{p}\right),e^{-r\Delta t}\sigma_{f}^{2}\sigma_{d_{n+1}-1}\sigma_{d_{n+1}}\sum_{q=1}^{P}\mathbf{\omega}_{q}h_{q}^{p}f_{q}^{p}\right)\label{eq:V_n}
\end{equation}
where $\sigma_{d_{n+1}-1}$, $\sigma_{d_{n+1}}$, $\sigma_{f}$ and
$\omega_{1},\dots,\omega_{P}$ are certain constants determined by
the GPR approximation of the function $\log\left(\mathbf{SV}_{1:n+1}\right)\mapsto v\left(t_{n+1},\mathbf{SV}_{1:n+1}\right)$
considering $\left\{ \mathbf{SV}_{\max\left\{ 1,n+1-J\right\} :n+1}^{p},p=1,\dots,P\right\} $
as the predictor set. Moreover, $h_{q}^{p}$ and $f_{q}^{p}$ are
two factors given by
\begin{equation}
h_{q}^{p}=\begin{cases}
\exp\left(-\sum_{i=1}^{d_{n+1}-2}\frac{\left(z_{i}^{p}-z_{i}^{q}\right)^{2}}{2\sigma_{i}^{2}}\right) & \text{if }n>0\\
1 & \text{if }n=0
\end{cases}
\end{equation}
and
\begin{equation}
f_{q}^{p}=\frac{\exp\left(-\frac{1}{2}\left(\left(\begin{array}{c}
z_{d_{n+1}-1}^{q}\\
z_{d_{n+1}}^{q}
\end{array}\right)-\boldsymbol{\mu}_{n+1,p}\right)^{\top}\left(\Sigma_{n+1,p}+\left(\begin{array}{cc}
\sigma_{d_{n+1}-1}^{2} & 0\\
0 & \sigma_{d_{n+1}}^{2}
\end{array}\right)\right)^{-1}\left(\left(\begin{array}{c}
z_{d_{n+1}-1}^{q}\\
z_{d_{n+1}}^{q}
\end{array}\right)-\boldsymbol{\mu}_{n+1,p}\right)\right)}{\sqrt{\text{\ensuremath{\det}}\left(\Sigma_{n+1,p}+\left(\begin{array}{cc}
\sigma_{d_{n+1}-1}^{2} & 0\\
0 & \sigma_{d_{n+1}}^{2}
\end{array}\right)\right)}},
\end{equation}
where $z_{i}^{p}=\log\left(S_{n+1-\left(i-1\right)/2}^{p}\right)$
if $i$ is even and $z_{i}^{p}=\log\left(V_{n+1-i/2}^{p}\right)$
if $i$ is odd, for $i=1,\dots,d_{n+1}$. 
\end{prop}

The proof of Proposition \ref{lem:L2} is reported in the Appendix
\ref{ApA3}. Relations (\ref{eq:v_Nm1}) and (\ref{eq:V_n}) can be
used to compute the option price at time $t=0$ by backward induction.

\section{Numerical results}

In this Section we present some numerical results about the effectiveness
of the proposed algorithms. The first section is devoted to the numerical
tests about the multi-dimensional Black-Scholes model, while the second
is devoted to the rough Bergomi model. The algorithms have been implemented
in MATLAB and computations have been preformed on a server which employs
a $2.40$ GHz Intel$^{{\scriptsize\textregistered}}$ Xenon$^{{\scriptsize\textregistered}}$
processor (Gold 6148, Skylake) and 20 GB of RAM. 

\subsection{Multi-dimensional Black-Scholes model}

Following Goudenège et al. \cite{goudenege2019machine}, we consider
an Arithmetic basket Put, a Geometric basket Put and a Call on the
Maximum of $d$-assets.

In particular, we use the following parameters $T=1$, $S_{0}^{i}=100$,
$K=100$, $r=0.05$, constant volatilities $\sigma_{i}=0.2$, constant
correlations $\rho_{ij}=0.2$ and $N=10$ exercise dates. Moreover,
we consider $P=250,\ 500$ or $1000$ points. As opposed to the other
input parameters, we vary the dimension $d$, considering $d=2,\,5,\,10,\,20,\,40$
and $100$. 

We present now the numerical results obtained with the GPR-Tree and
the GPR-EI methods for the three payoff examples.

\subsubsection{Geometric basket Put option}

Geometric basket Put is a particularly interesting option since it
is possible to reduce the problem of pricing it in the $d$-dimensional
model to a one dimensional American Put option in the Black-Scholes
model which can be priced straightforwardly, for example using the
CRR algorithm with $1000$ steps (see Cox et al. \cite{cox1979option}).
Therefore, in this case, we have a reliable benchmark to test the
proposed methods. Moreover, when $d$ is smaller than $10$ we can
also compute the price by means of a multi-dimensional binomial tree
(see Ekvall \cite{ekvall1996lattice}). In particular, the number
of steps employed for the multi-dimensional binomial tree is equal
to $200$ when $d=2$ and to $50$ when $d=5$. For values of $d$
larger than $5$, prices cannot be approximated via such a tree, because
the memory required for the calculations would be too large. Furthermore,
we also report the prices obtained with the GPR-MC method, employing
$P=1000$ points and $M=10^{5}$ Monte Carlo simulations, for comparison
purposes. As far as the GPR-Tree is concerned, we compute the prices
only for the values of $d$ smaller than $40$ since for higher values
of $d$ the tree step becomes over time demanding. In fact, the computation
of the continuation value with the tree step grows exponentially with
the dimension $d$ and for $d=40$ it requires the evaluation of the
GPR approximation at $2^{40}\approx10^{12}$ points for every times
step and for every point of $X$.

Results are reported in Table \ref{tab:GEO}. We observe that the
two proposed methods provide accurate and stable results and the computational
time is generally very small, except for the GPR-Tree method at $d=20$.
Moreover, the computer processing time of the GRP-EI method increases
little with the size of the problem and this makes the method particularly
effective when the dimension of the problem is high. This is because
the computation of the expected value and the training of the GPR
model are minimally affected by the dimension of the problem.

\begin{figure}
\begin{centering}
\includegraphics[width=1\textwidth]{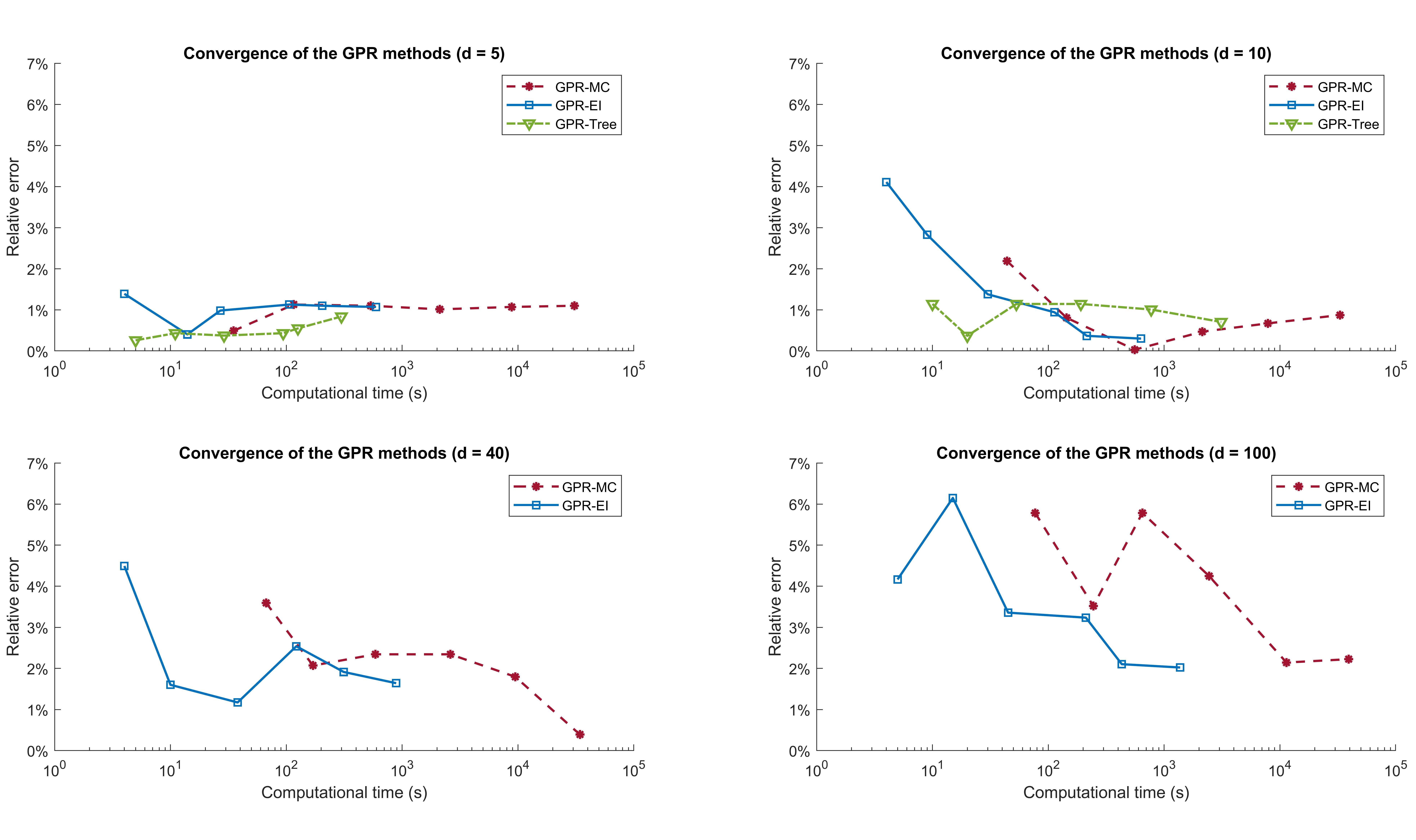}
\par\end{centering}
\caption{\label{fig:Comparison}{\small{}Comparison among the GPR methods changing
the dimension $d$ and doubling the number of points from $P=250$
to $P=8000$. As far as the GPR-MC method is concerned $M=10^{4}$
Monte Carlo simulations are employed.}}

\end{figure}

Figure \ref{fig:Comparison} investigates the convergence of the GPR
methods changing the dimension $d$. As we can see, the relative error
is small with all the considered methods, but the computational time
required by the GPR-Tree method and the GPR-EI method is generally
smaller with respect to the GPR-MC method.

\begin{table}
\begin{centering}
\begin{tabular}{cccccccccccccc}
\toprule 
 &  & \multicolumn{3}{c}{GPR-Tree} &  & \multicolumn{3}{c}{GPR-EI} &  & GPR-MC &  & Ekvall & Benchmark\tabularnewline
$d$ & $P$ & $\phantom{1}250$ & $\phantom{1}500$ & $1000$ &  & $\phantom{1}250$ & $\phantom{1}500$ & $1000$ &  &  &  &  & \tabularnewline
\midrule
2 &  & {\footnotesize{}$\underset{\left(4\right)}{4.61}$ } & {\footnotesize{}$\underset{\left(7\right)}{4.61}$ } & {\footnotesize{}$\underset{\left(22\right)}{4.61}$ } &  & {\footnotesize{}$\underset{\left(4\right)}{4.58}$} & {\footnotesize{}$\underset{\left(9\right)}{4.58}$} & {\footnotesize{}$\underset{\left(26\right)}{4.57}$} &  & {\footnotesize{}$4.57$} &  & {\footnotesize{}$4.62$} & {\footnotesize{}$4.62$}\tabularnewline
5 &  & {\footnotesize{}$\underset{\left(9\right)}{3.44}$ } & {\footnotesize{}$\underset{\left(15\right)}{3.43}$ } & {\footnotesize{}$\underset{\left(23\right)}{3.44}$ } &  & {\footnotesize{}$\underset{\left(4\right)}{3.40}$} & {\footnotesize{}$\underset{\left(14\right)}{3.43}$} & {\footnotesize{}$\underset{\left(27\right)}{3.41}$} &  & {\footnotesize{}$3.41$} &  & {\footnotesize{}$3.44$} & {\footnotesize{}$3.45$}\tabularnewline
10 &  & {\footnotesize{}$\underset{\left(10\right)}{3.00}$ } & {\footnotesize{}$\underset{\left(33\right)}{2.96}$} & {\footnotesize{}$\underset{\left(60\right)}{2.93}$ } &  & {\footnotesize{}$\underset{\left(4\right)}{2.85}$} & {\footnotesize{}$\underset{\left(9\right)}{2.88}$} & {\footnotesize{}$\underset{\left(30\right)}{2.93}$} &  & {\footnotesize{}$2.90$} &  &  & {\footnotesize{}$2.97$}\tabularnewline
20 &  & {\footnotesize{}$\underset{\left(4220\right)}{2.80}$ } & {\footnotesize{}$\underset{\left(14304\right)}{2.72}$} & {\footnotesize{}$\underset{\left(49609\right)}{2.72}$} &  & {\footnotesize{}$\underset{\left(4\right)}{2.63}$} & {\footnotesize{}$\underset{\left(9\right)}{2.73}$} & {\footnotesize{}$\underset{\left(29\right)}{2.63}$} &  & {\footnotesize{}$2.70$} &  &  & {\footnotesize{}$2.70$}\tabularnewline
40 &  &  &  &  &  & {\footnotesize{}$\underset{\left(4\right)}{2.45}$} & {\footnotesize{}$\underset{\left(10\right)}{2.52}$} & {\footnotesize{}$\underset{\left(38\right)}{2.53}$} &  & {\footnotesize{}$2.57$} &  &  & {\footnotesize{}$2.56$}\tabularnewline
100 &  &  &  &  &  & {\footnotesize{}$\underset{\left(5\right)}{2.27}$} & {\footnotesize{}$\underset{\left(15\right)}{2.32}$} & {\footnotesize{}$\underset{\left(45\right)}{2.39}$} &  & {\footnotesize{}$2.40$} &  &  & {\footnotesize{}$2.47$}\tabularnewline
\bottomrule
\end{tabular}
\par\end{centering}
\caption{\label{tab:GEO}{\small{}Results for a Geometric basket Put option
using the GPR-Tree method and the GPR-EI method. In the last columns,
the prices obtained by using the GPR-MC method, the Ekvall multi-dimensional
tree and the exact benchmark ($d$ is the dimension and $P$ is the
number of points). Values in brackets are the computational times
(in seconds).}}
\end{table}

\FloatBarrier

\subsubsection{Arithmetic basket Put option}

As opposed to the Geometric basket Put option, in this case we have
no method to obtain a fully reliable benchmark. Therefore we only
consider the prices obtained by means of the GPR-MC method, employed
with $P=1000$ points and $M=10^{5}$ Monte Carlo simulations. Moreover,
for small values of $d$, a benchmark can be obtained by means of
a multi-dimensional tree method (see Boyle et al. \cite{boyle1989numerical}),
just as shown for the Geometric case. Results are reported in Table
\ref{tab:ARI}. Similarly to the Geometric basket Put, the prices
obtained are reliable and they do not change much with respect to
the number $P$ of points. As opposed to the GPR-Tree method, which
can not be applied for high values of $d$, the GPR-EI method requires
a small computational time for all the values concerned of $d$.

\begin{table}[h]
\begin{centering}
\begin{tabular}{cccccccccccccc}
\toprule 
 &  & \multicolumn{3}{c}{GPR-Tree} &  & \multicolumn{3}{c}{GPR-EI} &  & GPR-MC &  & Ekvall & $\phantom{Benchmark}$\tabularnewline
$d$ & $P$ & $\phantom{1}250$ & $\phantom{1}500$ & $1000$ &  & $\phantom{1}250$ & $\phantom{1}500$ & $1000$ &  &  &  &  & \tabularnewline
\midrule
2 &  & {\footnotesize{}$\underset{\left(5\right)}{4.42}$} & {\footnotesize{}$\underset{\left(9\right)}{4.42}$} & {\footnotesize{}$\underset{\left(25\right)}{4.42}$} &  & {\footnotesize{}$\underset{\left(4\right)}{4.38}$} & {\footnotesize{}$\underset{\left(9\right)}{4.38}$} & {\footnotesize{}$\underset{\left(28\right)}{4.37}$} &  & {\footnotesize{}$4.37$} &  & {\footnotesize{}$4.42$} & \tabularnewline
5 &  & {\footnotesize{}$\underset{\left(5\right)}{3.15}$} & {\footnotesize{}$\underset{\left(9\right)}{3.12}$} & {\footnotesize{}$\underset{\left(24\right)}{3.13}$} &  & {\footnotesize{}$\underset{\left(6\right)}{3.09}$} & {\footnotesize{}$\underset{\left(9\right)}{3.12}$} & {\footnotesize{}$\underset{\left(44\right)}{3.10}$} &  & {\footnotesize{}$3.09$} &  & {\footnotesize{}$3.15$} & \tabularnewline
10 &  & {\footnotesize{}$\underset{\left(10\right)}{2.71}$} & {\footnotesize{}$\underset{\left(21\right)}{2.64}$} & {\footnotesize{}$\underset{\left(70\right)}{2.62}$} &  & {\footnotesize{}$\underset{\left(5\right)}{2.49}$} & {\footnotesize{}$\underset{\left(9\right)}{2.56}$} & {\footnotesize{}$\underset{\left(38\right)}{2.60}$} &  & {\footnotesize{}$2.58$} &  &  & \tabularnewline
20 &  & {\footnotesize{}$\underset{\left(4259\right)}{2.37}$} & {\footnotesize{}$\underset{\left(16343\right)}{2.35}$} & {\footnotesize{}$\underset{\left(57399\right)}{2.40}$} &  & {\footnotesize{}$\underset{\left(6\right)}{2.26}$} & {\footnotesize{}$\underset{\left(14\right)}{2.31}$} & {\footnotesize{}$\underset{\left(42\right)}{2.28}$} &  & {\footnotesize{}$2.38$} &  &  & \tabularnewline
40 &  &  &  &  &  & {\footnotesize{}$\underset{\left(4\right)}{2.18}$} & {\footnotesize{}$\underset{\left(10\right)}{2.18}$} & {\footnotesize{}$\underset{\left(31\right)}{2.16}$} &  & {\footnotesize{}$2.17$} &  &  & \tabularnewline
100 &  &  &  &  &  & {\footnotesize{}$\underset{\left(7\right)}{2.35}$} & {\footnotesize{}$\underset{\left(13\right)}{2.01}$} & {\footnotesize{}$\underset{\left(42\right)}{2.06}$} &  & {\footnotesize{}$1.92$} &  &  & \tabularnewline
\bottomrule
\end{tabular}
\par\end{centering}
\caption{\label{tab:ARI}{\small{}Results for an Arithmetic basket Put option
using the GPR-Tree method and the GPR-EI method. In the last columns,
the prices obtained by using the GPR-MC method and the Ekvall multi-dimensional
tree ($d$ is the dimension and $P$ is the number of points). Values
in brackets are the computational times (in seconds).}}
\end{table}

\subsubsection{Call on the Maximum}

As for the Arithmetic basket Put, in this case we have no numerical
methods to obtain a fully reliable benchmark. However, for small values
of $d$, we can approximate the price obtained by means of a multi-dimensional
tree method. Moreover, we also consider the price obtained with the
GPR-MC method. Results, which are shown in Table \ref{tab:MAX}, have
an accuracy comparable to the one obtained for the Arithmetic basket
Put option. 

\begin{table}[h]
\begin{centering}
\begin{tabular}{cccccccccccccc}
\toprule 
 &  & \multicolumn{3}{c}{GPR-Tree} &  & \multicolumn{3}{c}{GPR-EI} &  & GPR-MC &  & Ekvall & $\phantom{Benchmark}$\tabularnewline
$d$ & $P$ & $\phantom{1}250$ & $\phantom{1}500$ & $1000$ &  & $\phantom{1}250$ & $\phantom{1}500$ & $1000$ &  &  &  &  & \tabularnewline
\midrule
2 &  & {\footnotesize{}$\underset{\left(5\right)}{16.94}$ } & {\footnotesize{}$\underset{\left(8\right)}{16.94}$ } & {\footnotesize{}$\underset{\left(20\right)}{16.93}$ } &  & {\footnotesize{}$\underset{\left(4\right)}{16.75}$} & {\footnotesize{}$\underset{\left(10\right)}{16.81}$} & {\footnotesize{}$\underset{\left(28\right)}{16.82}$} &  & {\footnotesize{}$16.86$} &  & {\footnotesize{}$16.86$} & \tabularnewline
5 &  & {\footnotesize{}$\underset{\left(5\right)}{27.14}$ } & {\footnotesize{}$\underset{\left(10\right)}{27.17}$} & {\footnotesize{}$\underset{\left(26\right)}{27.19}$ } &  & {\footnotesize{}$\underset{\left(4\right)}{26.92}$} & {\footnotesize{}$\underset{\left(9\right)}{27.15}$} & {\footnotesize{}$\underset{\left(27\right)}{26.95}$} &  & {\footnotesize{}$27.20$} &  & {\footnotesize{}$27.20$} & \tabularnewline
10 &  & {\footnotesize{}$\underset{\left(11\right)}{35.27}$ } & {\footnotesize{}$\underset{\left(21\right)}{34.97}$ } & {\footnotesize{}$\underset{\left(106\right)}{35.08}$ } &  & {\footnotesize{}$\underset{\left(4\right)}{35.66}$} & {\footnotesize{}$\underset{\left(10\right)}{34.98}$} & {\footnotesize{}$\underset{\left(29\right)}{34.84}$} &  & {\footnotesize{}$35.17$} &  &  & \tabularnewline
20 &  & {\footnotesize{}$\underset{\left(4126\right)}{43.26}$} & {\footnotesize{}$\underset{\left(15025\right)}{43.21}$} & {\footnotesize{}$\underset{\left(51090\right)}{43.00}$} &  & {\footnotesize{}$\underset{\left(4\right)}{45.05}$} & {\footnotesize{}$\underset{\left(11\right)}{42.74}$} & {\footnotesize{}$\underset{\left(35\right)}{42.62}$} &  & {\footnotesize{}$42.76$} &  &  & \tabularnewline
40 &  &  &  &  &  & {\footnotesize{}$\underset{\left(5\right)}{51.79}$} & {\footnotesize{}$\underset{\left(10\right)}{50.36}$} & {\footnotesize{}$\underset{\left(41\right)}{49.53}$} &  & {\footnotesize{}$50.70$} &  &  & \tabularnewline
100 &  &  &  &  &  & {\footnotesize{}$\underset{\left(5\right)}{59.03}$} & {\footnotesize{}$\underset{\left(13\right)}{60.72}$} & {\footnotesize{}$\underset{\left(42\right)}{60.96}$} &  & {\footnotesize{}$59.69$} &  &  & \tabularnewline
\bottomrule
\end{tabular}
\par\end{centering}
\caption{\label{tab:MAX}{\small{}Results for a Call on Maximum Put option
using the GPR-Tree method and the GPR-EI method. In the last columns,
the prices obtained by using the GPR-MC method and the Ekvall multi-dimensional
tree ($d$ is the dimension and $P$ is the number of points). Values
in brackets are the computational times (in seconds).}}
\end{table}

\FloatBarrier

\subsection{Rough Bergomi model}

Following Bayer et al. \cite{bayer2018pricing}, we consider an American
Put option and we use the same parameters: $T=1$, $H=0.07$, $\rho=-0.90$,
$\xi_{0}=0.09$, $\eta=1.9$, $S_{0}=100$, $r=0.05$ and strike $K=70,80,\dots,120,130$
or $140$. As far as the GPR-Tree is concerned, we employ $N=50$
or $N=100$ time steps with $m=2$, $P=500,1000,2000$ or $4000$
random paths, and $J=0,1,3,7$ or $15$ past values. As far as the
GPR-EI is concerned, we employ $N=50$ or $N=100$ time steps, $P=1000,2000,4000$
or $8000$ random paths, and $J=0,1,3,7$ or $15$ past values. Similar
to what observed by Bayer et al. \cite{bayer2018pricing}, the difference
changing the value of $J$ does not impact significantly on the price,
which indicates that considering the non-Markovian nature of the processes
in the formulation of the exercise strategies is not particularly
relevant. Conversely, using a large number of predictors significantly
increases computational time. Numerical results are reported in Tables
\ref{tab:RB-Tree} and \ref{tab:RB-EI}, together with the results
reported by Bayer et al. in \cite{bayer2018pricing}. Prices are very
close to the benchmark, except for the case $K=120$: in this case
with both the two GPR methods we obtain a price which is close to
$20.20$ while Bayer et al. obtain $20.00$. Anyway, it is worth noticing
that the relative gap between these two results is less than $1\%$
.

\begin{table}[H]
\begin{centering}
\setlength\tabcolsep{6pt} \def\arraystretch{0.8} %
\begin{tabular}{ccccccccccccccc}
\toprule 
 &  &  &  & \multicolumn{9}{c}{{\small{}GPR-Tree}} &  & Bayer et al.\tabularnewline
\midrule 
 &  &  & {\small{}$N$} & \multicolumn{4}{c}{{\small{}$50$ }} &  & \multicolumn{4}{c}{{\small{}$100$}} &  & \tabularnewline
\cmidrule{4-15} 
{\small{}$K$} &  & {\small{}$J$} & {\small{}$P$} & {\small{}$\phantom{1}500$} & {\small{}$1000$} & {\small{}$2000$} & {\small{}$4000$} &  & {\small{}$\phantom{1}500$} & {\small{}$1000$} & {\small{}$2000$} & {\small{}$4000$} &  & \tabularnewline
\midrule 
{\small{}$70$} &  & {\scriptsize{}$0$} &  & {\footnotesize{}$\underset{\left(28\right)}{1.87}$} & {\footnotesize{}$\underset{\left(97\right)}{1.88}$} & {\footnotesize{}$\underset{\left(391\right)}{1.88}$} & {\footnotesize{}$\underset{\left(876\right)}{1.86}$} &  & {\footnotesize{}$\underset{\left(71\right)}{1.87}$} & {\footnotesize{}$\underset{\left(236\right)}{1.86}$} & {\footnotesize{}$\underset{\left(646\right)}{1.86}$} & {\footnotesize{}$\underset{\left(1337\right)}{1.86}$} &  & \multirow{5}{*}{{\scriptsize{}1.88}}\tabularnewline
 &  & {\scriptsize{}$1$} &  & {\footnotesize{}$\underset{\left(44\right)}{1.86}$} & {\footnotesize{}$\underset{\left(183\right)}{1.87}$} & {\footnotesize{}$\underset{\left(607\right)}{1.88}$} & {\footnotesize{}$\underset{\left(1672\right)}{1.87}$} &  & {\footnotesize{}$\underset{\left(95\right)}{1.86}$} & {\footnotesize{}$\underset{\left(310\right)}{1.87}$} & {\footnotesize{}$\underset{\left(1222\right)}{1.87}$} & {\footnotesize{}$\underset{\left(2265\right)}{1.87}$} &  & \tabularnewline
 &  & {\scriptsize{}$3$} &  & {\footnotesize{}$\underset{\left(71\right)}{1.87}$} & {\footnotesize{}$\underset{\left(296\right)}{1.86}$} & {\footnotesize{}$\underset{\left(1084\right)}{1.87}$} & {\footnotesize{}$\underset{\left(3742\right)}{1.87}$} &  & {\footnotesize{}$\underset{\left(163\right)}{1.86}$} & {\footnotesize{}$\underset{\left(594\right)}{1.87}$} & {\footnotesize{}$\underset{\left(1962\right)}{1.88}$} & {\footnotesize{}$\underset{\left(4222\right)}{1.88}$} &  & \tabularnewline
 &  & {\scriptsize{}$7$} &  & {\footnotesize{}$\underset{\left(168\right)}{1.91}$} & {\footnotesize{}$\underset{\left(563\right)}{1.87}$} & {\footnotesize{}$\underset{\left(1930\right)}{1.86}$} & {\footnotesize{}$\underset{\left(3501\right)}{1.87}$} &  & {\footnotesize{}$\underset{\left(275\right)}{1.85}$} & {\footnotesize{}$\underset{\left(1141\right)}{1.87}$} & {\footnotesize{}$\underset{\left(4579\right)}{1.88}$} & {\footnotesize{}$\underset{\left(7997\right)}{1.88}$} &  & \tabularnewline
 &  & {\scriptsize{}$15$} &  & {\footnotesize{}$\underset{\left(248\right)}{1.94}$} & {\footnotesize{}$\underset{\left(986\right)}{1.88}$} & {\footnotesize{}$\underset{\left(4841\right)}{1.87}$} & {\footnotesize{}$\underset{\left(7806\right)}{1.87}$} &  & {\footnotesize{}$\underset{\left(541\right)}{1.87}$} & {\footnotesize{}$\underset{\left(2171\right)}{1.88}$} & {\footnotesize{}$\underset{\left(10169\right)}{1.87}$} & {\footnotesize{}$\underset{\left(16466\right)}{1.86}$} &  & \tabularnewline
\midrule
{\small{}$80$} &  & {\scriptsize{}$0$} &  & {\footnotesize{}$\underset{\left(31\right)}{3.18}$} & {\footnotesize{}$\underset{\left(117\right)}{3.19}$} & {\footnotesize{}$\underset{\left(376\right)}{3.20}$} & {\footnotesize{}$\underset{\left(823\right)}{3.20}$} &  & {\footnotesize{}$\underset{\left(85\right)}{3.17}$} & {\footnotesize{}$\underset{\left(216\right)}{3.18}$} & {\footnotesize{}$\underset{\left(603\right)}{3.19}$} & {\footnotesize{}$\underset{\left(1368\right)}{3.19}$} &  & \multirow{5}{*}{{\scriptsize{}3.22}}\tabularnewline
 &  & {\scriptsize{}$1$} &  & {\footnotesize{}$\underset{\left(47\right)}{3.19}$} & {\footnotesize{}$\underset{\left(152\right)}{3.19}$} & {\footnotesize{}$\underset{\left(569\right)}{3.20}$} & {\footnotesize{}$\underset{\left(1166\right)}{3.20}$} &  & {\footnotesize{}$\underset{\left(119\right)}{3.18}$} & {\footnotesize{}$\underset{\left(322\right)}{3.20}$} & {\footnotesize{}$\underset{\left(1107\right)}{3.20}$} & {\footnotesize{}$\underset{\left(2396\right)}{3.20}$} &  & \tabularnewline
 &  & {\scriptsize{}$3$} &  & {\footnotesize{}$\underset{\left(93\right)}{3.19}$} & {\footnotesize{}$\underset{\left(287\right)}{3.19}$} & {\footnotesize{}$\underset{\left(1070\right)}{3.20}$} & {\footnotesize{}$\underset{\left(2095\right)}{3.20}$} &  & {\footnotesize{}$\underset{\left(167\right)}{3.17}$} & {\footnotesize{}$\underset{\left(617\right)}{3.21}$} & {\footnotesize{}$\underset{\left(1966\right)}{3.21}$} & {\footnotesize{}$\underset{\left(4043\right)}{3.22}$} &  & \tabularnewline
 &  & {\scriptsize{}$7$} &  & {\footnotesize{}$\underset{\left(136\right)}{3.21}$} & {\footnotesize{}$\underset{\left(624\right)}{3.20}$} & {\footnotesize{}$\underset{\left(2653\right)}{3.20}$} & {\footnotesize{}$\underset{\left(3721\right)}{3.21}$} &  & {\footnotesize{}$\underset{\left(301\right)}{3.19}$} & {\footnotesize{}$\underset{\left(1134\right)}{3.21}$} & {\footnotesize{}$\underset{\left(4179\right)}{3.21}$} & {\footnotesize{}$\underset{\left(8031\right)}{3.23}$} &  & \tabularnewline
 &  & {\scriptsize{}$15$} &  & {\footnotesize{}$\underset{\left(322\right)}{3.24}$} & {\footnotesize{}$\underset{\left(1186\right)}{3.20}$} & {\footnotesize{}$\underset{\left(7011\right)}{3.20}$} & {\footnotesize{}$\underset{\left(7392\right)}{3.21}$} &  & {\footnotesize{}$\underset{\left(633\right)}{3.18}$} & {\footnotesize{}$\underset{\left(1940\right)}{3.20}$} & {\footnotesize{}$\underset{\left(10317\right)}{3.20}$} & {\footnotesize{}$\underset{\left(20584\right)}{3.23}$} &  & \tabularnewline
\midrule
{\small{}$90$} &  & {\scriptsize{}$0$} &  & {\footnotesize{}$\underset{\left(28\right)}{5.24}$} & {\footnotesize{}$\underset{\left(102\right)}{5.24}$} & {\footnotesize{}$\underset{\left(359\right)}{5.25}$} & {\footnotesize{}$\underset{\left(702\right)}{5.26}$} &  & {\footnotesize{}$\underset{\left(82\right)}{5.25}$} & {\footnotesize{}$\underset{\left(223\right)}{5.28}$} & {\footnotesize{}$\underset{\left(707\right)}{5.26}$} & {\footnotesize{}$\underset{\left(1504\right)}{5.28}$} &  & \multirow{5}{*}{{\scriptsize{}5.31}}\tabularnewline
 &  & {\scriptsize{}$1$} &  & {\footnotesize{}$\underset{\left(44\right)}{5.25}$} & {\footnotesize{}$\underset{\left(163\right)}{5.25}$} & {\footnotesize{}$\underset{\left(512\right)}{5.26}$} & {\footnotesize{}$\underset{\left(1185\right)}{5.27}$} &  & {\footnotesize{}$\underset{\left(109\right)}{5.27}$} & {\footnotesize{}$\underset{\left(283\right)}{5.29}$} & {\footnotesize{}$\underset{\left(1144\right)}{5.28}$} & {\footnotesize{}$\underset{\left(3954\right)}{5.30}$} &  & \tabularnewline
 &  & {\scriptsize{}$3$} &  & {\footnotesize{}$\underset{\left(94\right)}{5.27}$} & {\footnotesize{}$\underset{\left(330\right)}{5.26}$} & {\footnotesize{}$\underset{\left(1058\right)}{5.28}$} & {\footnotesize{}$\underset{\left(1756\right)}{5.28}$} &  & {\footnotesize{}$\underset{\left(177\right)}{5.30}$} & {\footnotesize{}$\underset{\left(555\right)}{5.31}$} & {\footnotesize{}$\underset{\left(1833\right)}{5.31}$} & {\footnotesize{}$\underset{\left(4226\right)}{5.32}$} &  & \tabularnewline
 &  & {\scriptsize{}$7$} &  & {\footnotesize{}$\underset{\left(150\right)}{5.28}$} & {\footnotesize{}$\underset{\left(561\right)}{5.30}$} & {\footnotesize{}$\underset{\left(2253\right)}{5.29}$} & {\footnotesize{}$\underset{\left(3595\right)}{5.29}$} &  & {\footnotesize{}$\underset{\left(315\right)}{5.30}$} & {\footnotesize{}$\underset{\left(1089\right)}{5.33}$} & {\footnotesize{}$\underset{\left(4319\right)}{5.33}$} & {\footnotesize{}$\underset{\left(7073\right)}{5.33}$} &  & \tabularnewline
 &  & {\scriptsize{}$15$} &  & {\footnotesize{}$\underset{\left(269\right)}{5.28}$} & {\footnotesize{}$\underset{\left(1000\right)}{5.27}$} & {\footnotesize{}$\underset{\left(4348\right)}{5.28}$} & {\footnotesize{}$\underset{\left(7411\right)}{5.29}$} &  & {\footnotesize{}$\underset{\left(533\right)}{5.28}$} & {\footnotesize{}$\underset{\left(2127\right)}{5.33}$} & {\footnotesize{}$\underset{\left(17098\right)}{5.34}$} & {\footnotesize{}$\underset{\left(16804\right)}{5.34}$} &  & \tabularnewline
\midrule
{\small{}$100$} &  & {\scriptsize{}$0$} &  & {\footnotesize{}$\underset{\left(29\right)}{8.36}$} & {\footnotesize{}$\underset{\left(103\right)}{8.37}$} & {\footnotesize{}$\underset{\left(329\right)}{8.37}$} & {\footnotesize{}$\underset{\left(748\right)}{8.39}$} &  & {\footnotesize{}$\underset{\left(70\right)}{8.42}$} & {\footnotesize{}$\underset{\left(190\right)}{8.45}$} & {\footnotesize{}$\underset{\left(584\right)}{8.42}$} & {\footnotesize{}$\underset{\left(1313\right)}{8.46}$} &  & \multirow{5}{*}{{\scriptsize{}8.50}}\tabularnewline
 &  & {\scriptsize{}$1$} &  & {\footnotesize{}$\underset{\left(47\right)}{8.39}$} & {\footnotesize{}$\underset{\left(177\right)}{8.40}$} & {\footnotesize{}$\underset{\left(510\right)}{8.39}$} & {\footnotesize{}$\underset{\left(1145\right)}{8.42}$} &  & {\footnotesize{}$\underset{\left(89\right)}{8.43}$} & {\footnotesize{}$\underset{\left(325\right)}{8.46}$} & {\footnotesize{}$\underset{\left(969\right)}{8.44}$} & {\footnotesize{}$\underset{\left(2058\right)}{8.48}$} &  & \tabularnewline
 &  & {\scriptsize{}$3$} &  & {\footnotesize{}$\underset{\left(89\right)}{8.42}$} & {\footnotesize{}$\underset{\left(302\right)}{8.42}$} & {\footnotesize{}$\underset{\left(986\right)}{8.43}$} & {\footnotesize{}$\underset{\left(1844\right)}{8.45}$} &  & {\footnotesize{}$\underset{\left(167\right)}{8.47}$} & {\footnotesize{}$\underset{\left(551\right)}{8.50}$} & {\footnotesize{}$\underset{\left(2322\right)}{8.49}$} & {\footnotesize{}$\underset{\left(4439\right)}{8.51}$} &  & \tabularnewline
 &  & {\scriptsize{}$7$} &  & {\footnotesize{}$\underset{\left(173\right)}{8.43}$} & {\footnotesize{}$\underset{\left(552\right)}{8.43}$} & {\footnotesize{}$\underset{\left(2083\right)}{8.44}$} & {\footnotesize{}$\underset{\left(3926\right)}{8.45}$} &  & {\footnotesize{}$\underset{\left(322\right)}{8.47}$} & {\footnotesize{}$\underset{\left(1117\right)}{8.51}$} & {\footnotesize{}$\underset{\left(4120\right)}{8.49}$} & {\footnotesize{}$\underset{\left(8324\right)}{8.53}$} &  & \tabularnewline
 &  & {\scriptsize{}$15$} &  & {\footnotesize{}$\underset{\left(340\right)}{8.44}$} & {\footnotesize{}$\underset{\left(1134\right)}{8.44}$} & {\footnotesize{}$\underset{\left(4637\right)}{8.45}$} & {\footnotesize{}$\underset{\left(7013\right)}{8.46}$} &  & {\footnotesize{}$\underset{\left(684\right)}{8.51}$} & {\footnotesize{}$\underset{\left(2229\right)}{8.53}$} & {\footnotesize{}$\underset{\left(8403\right)}{8.48}$} & {\footnotesize{}$\underset{\left(14183\right)}{8.53}$} &  & \tabularnewline
\midrule
{\small{}$110$} &  & {\scriptsize{}$0$} &  & {\footnotesize{}$\underset{\left(32\right)}{13.04}$} & {\footnotesize{}$\underset{\left(90\right)}{13.06}$} & {\footnotesize{}$\underset{\left(334\right)}{13.08}$} & {\footnotesize{}$\underset{\left(695\right)}{13.12}$} &  & {\footnotesize{}$\underset{\left(77\right)}{13.15}$} & {\footnotesize{}$\underset{\left(237\right)}{13.18}$} & {\footnotesize{}$\underset{\left(572\right)}{13.16}$} & {\footnotesize{}$\underset{\left(1364\right)}{13.20}$} &  & \multirow{5}{*}{{\scriptsize{}13.23}}\tabularnewline
 &  & {\scriptsize{}$1$} &  & {\footnotesize{}$\underset{\left(67\right)}{13.09}$} & {\footnotesize{}$\underset{\left(180\right)}{13.09}$} & {\footnotesize{}$\underset{\left(544\right)}{13.12}$} & {\footnotesize{}$\underset{\left(1135\right)}{13.15}$} &  & {\footnotesize{}$\underset{\left(95\right)}{13.17}$} & {\footnotesize{}$\underset{\left(296\right)}{13.20}$} & {\footnotesize{}$\underset{\left(1192\right)}{13.19}$} & {\footnotesize{}$\underset{\left(2207\right)}{13.22}$} &  & \tabularnewline
 &  & {\scriptsize{}$3$} &  & {\footnotesize{}$\underset{\left(78\right)}{13.11}$} & {\footnotesize{}$\underset{\left(282\right)}{13.14}$} & {\footnotesize{}$\underset{\left(1119\right)}{13.17}$} & {\footnotesize{}$\underset{\left(1896\right)}{13.18}$} &  & {\footnotesize{}$\underset{\left(158\right)}{13.18}$} & {\footnotesize{}$\underset{\left(575\right)}{13.23}$} & {\footnotesize{}$\underset{\left(1917\right)}{13.23}$} & {\footnotesize{}$\underset{\left(4028\right)}{13.26}$} &  & \tabularnewline
 &  & {\scriptsize{}$7$} &  & {\footnotesize{}$\underset{\left(157\right)}{13.11}$} & {\footnotesize{}$\underset{\left(520\right)}{13.14}$} & {\footnotesize{}$\underset{\left(2083\right)}{13.19}$} & {\footnotesize{}$\underset{\left(3659\right)}{13.19}$} &  & {\footnotesize{}$\underset{\left(318\right)}{13.20}$} & {\footnotesize{}$\underset{\left(1058\right)}{13.22}$} & {\footnotesize{}$\underset{\left(4508\right)}{13.24}$} & {\footnotesize{}$\underset{\left(7440\right)}{13.29}$} &  & \tabularnewline
 &  & {\scriptsize{}$15$} &  & {\footnotesize{}$\underset{\left(254\right)}{13.09}$} & {\footnotesize{}$\underset{\left(1007\right)}{13.15}$} & {\footnotesize{}$\underset{\left(4449\right)}{13.17}$} & {\footnotesize{}$\underset{\left(7668\right)}{13.20}$} &  & {\footnotesize{}$\underset{\left(625\right)}{13.22}$} & {\footnotesize{}$\underset{\left(2582\right)}{13.26}$} & {\footnotesize{}$\underset{\left(9055\right)}{13.27}$} & {\footnotesize{}$\underset{\left(13191\right)}{13.24}$} &  & \tabularnewline
\midrule
{\small{}$120$} &  & {\scriptsize{}$0$} &  & {\footnotesize{}$\underset{\left(37\right)}{20.19}$} & {\footnotesize{}$\underset{\left(121\right)}{20.19}$} & {\footnotesize{}$\underset{\left(304\right)}{20.20}$} & {\footnotesize{}$\underset{\left(692\right)}{20.22}$} &  & {\footnotesize{}$\underset{\left(79\right)}{20.21}$} & {\footnotesize{}$\underset{\left(206\right)}{20.24}$} & {\footnotesize{}$\underset{\left(662\right)}{20.22}$} & {\footnotesize{}$\underset{\left(1484\right)}{20.23}$} &  & \multirow{5}{*}{{\scriptsize{}20.00}}\tabularnewline
 &  & {\scriptsize{}$1$} &  & {\footnotesize{}$\underset{\left(49\right)}{20.20}$} & {\footnotesize{}$\underset{\left(180\right)}{20.21}$} & {\footnotesize{}$\underset{\left(494\right)}{20.21}$} & {\footnotesize{}$\underset{\left(1047\right)}{20.25}$} &  & {\footnotesize{}$\underset{\left(95\right)}{20.21}$} & {\footnotesize{}$\underset{\left(283\right)}{20.24}$} & {\footnotesize{}$\underset{\left(959\right)}{20.24}$} & {\footnotesize{}$\underset{\left(2029\right)}{20.26}$} &  & \tabularnewline
 &  & {\scriptsize{}$3$} &  & {\footnotesize{}$\underset{\left(98\right)}{20.19}$} & {\footnotesize{}$\underset{\left(268\right)}{20.19}$} & {\footnotesize{}$\underset{\left(1120\right)}{20.25}$} & {\footnotesize{}$\underset{\left(2077\right)}{20.26}$} &  & {\footnotesize{}$\underset{\left(156\right)}{20.23}$} & {\footnotesize{}$\underset{\left(588\right)}{20.26}$} & {\footnotesize{}$\underset{\left(2075\right)}{20.26}$} & {\footnotesize{}$\underset{\left(3705\right)}{20.24}$} &  & \tabularnewline
 &  & {\scriptsize{}$7$} &  & {\footnotesize{}$\underset{\left(152\right)}{20.20}$} & {\footnotesize{}$\underset{\left(511\right)}{20.18}$} & {\footnotesize{}$\underset{\left(1935\right)}{20.17}$} & {\footnotesize{}$\underset{\left(3592\right)}{20.26}$} &  & {\footnotesize{}$\underset{\left(363\right)}{20.25}$} & {\footnotesize{}$\underset{\left(1139\right)}{20.25}$} & {\footnotesize{}$\underset{\left(4395\right)}{20.23}$} & {\footnotesize{}$\underset{\left(6293\right)}{20.28}$} &  & \tabularnewline
 &  & {\scriptsize{}$15$} &  & {\footnotesize{}$\underset{\left(278\right)}{20.19}$} & {\footnotesize{}$\underset{\left(1036\right)}{20.17}$} & {\footnotesize{}$\underset{\left(4161\right)}{20.22}$} & {\footnotesize{}$\underset{\left(7844\right)}{20.24}$} &  & {\footnotesize{}$\underset{\left(624\right)}{20.18}$} & {\footnotesize{}$\underset{\left(1951\right)}{20.22}$} & {\footnotesize{}$\underset{\left(8057\right)}{20.19}$} & {\footnotesize{}$\underset{\left(15643\right)}{20.28}$} &  & \tabularnewline
\midrule
{\small{}$130$} &  &  &  & \multicolumn{4}{c}{{\scriptsize{}Always $30.00$}} &  & \multicolumn{4}{c}{{\scriptsize{}Always $30.00$}} &  & {\scriptsize{}30.00}\tabularnewline
\midrule
{\small{}$140$} &  &  &  & \multicolumn{4}{c}{{\scriptsize{}Always $40.00$}} &  & \multicolumn{4}{c}{{\scriptsize{}Always $40.00$}} &  & {\scriptsize{}40.00}\tabularnewline
\bottomrule
\end{tabular}
\par\end{centering}
\caption{\label{tab:RB-Tree}{\small{}Results for an American Put option in
the rough Bergomi model using the GPR-Tree method. $N$ represents
the number of time steps, $P$ the number of the simulated paths and
$J$ the number of past values employed in the regression. Values
in brackets are the computational times (in seconds).}}
\end{table}

\begin{table}[H]
\begin{centering}
\setlength\tabcolsep{6pt} \def\arraystretch{0.8} %
\begin{tabular}{cccccccccccccc}
\toprule 
 &  &  & \multicolumn{9}{c}{{\small{}GPR-EI}} &  & Bayer et al.\tabularnewline
\midrule 
 &  & {\small{}$N$} & \multicolumn{4}{c}{{\small{}$50$ }} &  & \multicolumn{4}{c}{{\small{}$100$}} &  & \tabularnewline
\cmidrule{3-14} 
{\small{}$K$} & {\small{}$J$} & {\small{}$P$} & {\small{}$1000$} & {\small{}$2000$} & {\small{}$4000$} & {\small{}$8000$} &  & {\small{}$1000$} & {\small{}$2000$} & {\small{}$4000$} & {\small{}$8000$} &  & \tabularnewline
\midrule 
{\small{}$70$} & {\footnotesize{}$0$} &  & {\footnotesize{}$\underset{\left(101\right)}{1.82}$} & {\footnotesize{}$\underset{\left(253\right)}{1.84}$} & {\footnotesize{}$\underset{\left(351\right)}{1.85}$} & {\footnotesize{}$\underset{\left(533\right)}{1.85}$} &  & {\footnotesize{}$\underset{\left(162\right)}{1.86}$} & {\footnotesize{}$\underset{\left(579\right)}{1.88}$} & {\footnotesize{}$\underset{\left(689\right)}{1.87}$} & {\footnotesize{}$\underset{\left(1011\right)}{1.88}$} &  & \multirow{5}{*}{{\scriptsize{}1.88}}\tabularnewline
 & {\footnotesize{}$1$} &  & {\footnotesize{}$\underset{\left(96\right)}{1.82}$} & {\footnotesize{}$\underset{\left(525\right)}{1.85}$} & {\footnotesize{}$\underset{\left(636\right)}{1.85}$} & {\footnotesize{}$\underset{\left(884\right)}{1.85}$} &  & {\footnotesize{}$\underset{\left(184\right)}{1.86}$} & {\footnotesize{}$\underset{\left(816\right)}{1.88}$} & {\footnotesize{}$\underset{\left(913\right)}{1.87}$} & {\footnotesize{}$\underset{\left(1551\right)}{1.88}$} &  & \tabularnewline
 & {\footnotesize{}$3$} &  & {\footnotesize{}$\underset{\left(263\right)}{1.83}$} & {\footnotesize{}$\underset{\left(1305\right)}{1.85}$} & {\footnotesize{}$\underset{\left(1118\right)}{1.83}$} & {\footnotesize{}$\underset{\left(1630\right)}{1.84}$} &  & {\footnotesize{}$\underset{\left(369\right)}{1.86}$} & {\footnotesize{}$\underset{\left(2389\right)}{1.88}$} & {\footnotesize{}$\underset{\left(2831\right)}{1.88}$} & {\footnotesize{}$\underset{\left(2994\right)}{1.89}$} &  & \tabularnewline
 & {\footnotesize{}$7$} &  & {\footnotesize{}$\underset{\left(497\right)}{1.81}$} & {\footnotesize{}$\underset{\left(2706\right)}{1.85}$} & {\footnotesize{}$\underset{\left(3014\right)}{1.85}$} & {\footnotesize{}$\underset{\left(3447\right)}{1.85}$} &  & {\footnotesize{}$\underset{\left(657\right)}{1.80}$} & {\footnotesize{}$\underset{\left(4848\right)}{1.87}$} & {\footnotesize{}$\underset{\left(5576\right)}{1.88}$} & {\footnotesize{}$\underset{\left(4132\right)}{1.86}$} &  & \tabularnewline
 & {\footnotesize{}$15$} &  & {\footnotesize{}$\underset{\left(820\right)}{1.78}$} & {\footnotesize{}$\underset{\left(4939\right)}{1.84}$} & {\footnotesize{}$\underset{\left(5802\right)}{1.83}$} & {\footnotesize{}$\underset{\left(6006\right)}{1.83}$} &  & {\footnotesize{}$\underset{\left(1932\right)}{1.79}$} & {\footnotesize{}$\underset{\left(11876\right)}{1.83}$} & {\footnotesize{}$\underset{\left(14703\right)}{1.85}$} & {\footnotesize{}$\underset{\left(5870\right)}{1.88}$} &  & \tabularnewline
\midrule
{\small{}$80$} & {\footnotesize{}$0$} &  & {\footnotesize{}$\underset{\left(86\right)}{3.14}$} & {\footnotesize{}$\underset{\left(271\right)}{3.16}$} & {\footnotesize{}$\underset{\left(348\right)}{3.18}$} & {\footnotesize{}$\underset{\left(558\right)}{3.17}$} &  & {\footnotesize{}$\underset{\left(162\right)}{3.22}$} & {\footnotesize{}$\underset{\left(549\right)}{3.24}$} & {\footnotesize{}$\underset{\left(602\right)}{3.21}$} & {\footnotesize{}$\underset{\left(1065\right)}{3.22}$} &  & \multirow{5}{*}{{\scriptsize{}3.22}}\tabularnewline
 & {\footnotesize{}$1$} &  & {\footnotesize{}$\underset{\left(127\right)}{3.14}$} & {\footnotesize{}$\underset{\left(409\right)}{3.16}$} & {\footnotesize{}$\underset{\left(601\right)}{3.19}$} & {\footnotesize{}$\underset{\left(865\right)}{3.18}$} &  & {\footnotesize{}$\underset{\left(212\right)}{3.23}$} & {\footnotesize{}$\underset{\left(984\right)}{3.24}$} & {\footnotesize{}$\underset{\left(847\right)}{3.21}$} & {\footnotesize{}$\underset{\left(1285\right)}{3.22}$} &  & \tabularnewline
 & {\footnotesize{}$3$} &  & {\footnotesize{}$\underset{\left(160\right)}{3.14}$} & {\footnotesize{}$\underset{\left(1334\right)}{3.18}$} & {\footnotesize{}$\underset{\left(1190\right)}{3.19}$} & {\footnotesize{}$\underset{\left(1476\right)}{3.19}$} &  & {\footnotesize{}$\underset{\left(357\right)}{3.22}$} & {\footnotesize{}$\underset{\left(1411\right)}{3.24}$} & {\footnotesize{}$\underset{\left(2739\right)}{3.23}$} & {\footnotesize{}$\underset{\left(2387\right)}{3.21}$} &  & \tabularnewline
 & {\footnotesize{}$7$} &  & {\footnotesize{}$\underset{\left(453\right)}{3.15}$} & {\footnotesize{}$\underset{\left(3263\right)}{3.18}$} & {\footnotesize{}$\underset{\left(3197\right)}{3.19}$} & {\footnotesize{}$\underset{\left(3252\right)}{3.18}$} &  & {\footnotesize{}$\underset{\left(631\right)}{3.22}$} & {\footnotesize{}$\underset{\left(5813\right)}{3.24}$} & {\footnotesize{}$\underset{\left(5327\right)}{3.23}$} & {\footnotesize{}$\underset{\left(5035\right)}{3.25}$} &  & \tabularnewline
 & {\footnotesize{}$15$} &  & {\footnotesize{}$\underset{\left(947\right)}{3.12}$} & {\footnotesize{}$\underset{\left(7107\right)}{3.16}$} & {\footnotesize{}$\underset{\left(5650\right)}{3.19}$} & {\footnotesize{}$\underset{\left(7575\right)}{3.16}$} &  & {\footnotesize{}$\underset{\left(2103\right)}{3.17}$} & {\footnotesize{}$\underset{\left(17466\right)}{3.12}$} & {\footnotesize{}$\underset{\left(15258\right)}{3.23}$} & {\footnotesize{}$\underset{\left(5974\right)}{3.22}$} &  & \tabularnewline
\midrule
{\small{}$90$} & {\footnotesize{}$0$} &  & {\footnotesize{}$\underset{\left(77\right)}{5.19}$} & {\footnotesize{}$\underset{\left(271\right)}{5.22}$} & {\footnotesize{}$\underset{\left(353\right)}{5.24}$} & {\footnotesize{}$\underset{\left(517\right)}{5.24}$} &  & {\footnotesize{}$\underset{\left(166\right)}{5.29}$} & {\footnotesize{}$\underset{\left(470\right)}{5.30}$} & {\footnotesize{}$\underset{\left(608\right)}{5.28}$} & {\footnotesize{}$\underset{\left(993\right)}{5.29}$} &  & \multirow{5}{*}{{\scriptsize{}5.31}}\tabularnewline
 & {\footnotesize{}$1$} &  & {\footnotesize{}$\underset{\left(89\right)}{5.19}$} & {\footnotesize{}$\underset{\left(416\right)}{5.22}$} & {\footnotesize{}$\underset{\left(455\right)}{5.24}$} & {\footnotesize{}$\underset{\left(748\right)}{5.25}$} &  & {\footnotesize{}$\underset{\left(223\right)}{5.31}$} & {\footnotesize{}$\underset{\left(887\right)}{5.32}$} & {\footnotesize{}$\underset{\left(1146\right)}{5.30}$} & {\footnotesize{}$\underset{\left(1266\right)}{5.29}$} &  & \tabularnewline
 & {\footnotesize{}$3$} &  & {\footnotesize{}$\underset{\left(239\right)}{5.22}$} & {\footnotesize{}$\underset{\left(1036\right)}{5.26}$} & {\footnotesize{}$\underset{\left(1259\right)}{5.27}$} & {\footnotesize{}$\underset{\left(1230\right)}{5.24}$} &  & {\footnotesize{}$\underset{\left(493\right)}{5.33}$} & {\footnotesize{}$\underset{\left(2624\right)}{5.34}$} & {\footnotesize{}$\underset{\left(1427\right)}{5.28}$} & {\footnotesize{}$\underset{\left(2387\right)}{5.33}$} &  & \tabularnewline
 & {\footnotesize{}$7$} &  & {\footnotesize{}$\underset{\left(307\right)}{5.19}$} & {\footnotesize{}$\underset{\left(2490\right)}{5.23}$} & {\footnotesize{}$\underset{\left(2348\right)}{5.26}$} & {\footnotesize{}$\underset{\left(2534\right)}{5.25}$} &  & {\footnotesize{}$\underset{\left(1584\right)}{5.32}$} & {\footnotesize{}$\underset{\left(3909\right)}{5.30}$} & {\footnotesize{}$\underset{\left(4560\right)}{5.30}$} & {\footnotesize{}$\underset{\left(5803\right)}{5.34}$} &  & \tabularnewline
 & {\footnotesize{}$15$} &  & {\footnotesize{}$\underset{\left(1189\right)}{5.23}$} & {\footnotesize{}$\underset{\left(5729\right)}{5.25}$} & {\footnotesize{}$\underset{\left(6236\right)}{5.26}$} & {\footnotesize{}$\underset{\left(6503\right)}{5.27}$} &  & {\footnotesize{}$\underset{\left(2120\right)}{5.28}$} & {\footnotesize{}$\underset{\left(9220\right)}{5.28}$} & {\footnotesize{}$\underset{\left(9943\right)}{5.28}$} & {\footnotesize{}$\underset{\left(6216\right)}{5.29}$} &  & \tabularnewline
\midrule
{\small{}$100$} & {\footnotesize{}$0$} &  & {\footnotesize{}$\underset{\left(81\right)}{8.30}$} & {\footnotesize{}$\underset{\left(260\right)}{8.33}$} & {\footnotesize{}$\underset{\left(472\right)}{8.36}$} & {\footnotesize{}$\underset{\left(566\right)}{8.38}$} &  & {\footnotesize{}$\underset{\left(189\right)}{8.44}$} & {\footnotesize{}$\underset{\left(466\right)}{8.46}$} & {\footnotesize{}$\underset{\left(625\right)}{8.45}$} & {\footnotesize{}$\underset{\left(1099\right)}{8.45}$} &  & \multirow{5}{*}{{\scriptsize{}8.50}}\tabularnewline
 & {\footnotesize{}$1$} &  & {\footnotesize{}$\underset{\left(93\right)}{8.30}$} & {\footnotesize{}$\underset{\left(402\right)}{8.33}$} & {\footnotesize{}$\underset{\left(413\right)}{8.36}$} & {\footnotesize{}$\underset{\left(732\right)}{8.38}$} &  & {\footnotesize{}$\underset{\left(191\right)}{8.44}$} & {\footnotesize{}$\underset{\left(742\right)}{8.46}$} & {\footnotesize{}$\underset{\left(1189\right)}{8.48}$} & {\footnotesize{}$\underset{\left(1362\right)}{8.46}$} &  & \tabularnewline
 & {\footnotesize{}$3$} &  & {\footnotesize{}$\underset{\left(250\right)}{8.37}$} & {\footnotesize{}$\underset{\left(851\right)}{8.35}$} & {\footnotesize{}$\underset{\left(1412\right)}{8.43}$} & {\footnotesize{}$\underset{\left(1028\right)}{8.38}$} &  & {\footnotesize{}$\underset{\left(362\right)}{8.44}$} & {\footnotesize{}$\underset{\left(1256\right)}{8.46}$} & {\footnotesize{}$\underset{\left(2344\right)}{8.51}$} & {\footnotesize{}$\underset{\left(1886\right)}{8.45}$} &  & \tabularnewline
 & {\footnotesize{}$7$} &  & {\footnotesize{}$\underset{\left(476\right)}{8.39}$} & {\footnotesize{}$\underset{\left(2957\right)}{8.39}$} & {\footnotesize{}$\underset{\left(3366\right)}{8.44}$} & {\footnotesize{}$\underset{\left(3556\right)}{8.42}$} &  & {\footnotesize{}$\underset{\left(670\right)}{8.44}$} & {\footnotesize{}$\underset{\left(3808\right)}{8.47}$} & {\footnotesize{}$\underset{\left(4867\right)}{8.53}$} & {\footnotesize{}$\underset{\left(5165\right)}{8.52}$} &  & \tabularnewline
 & {\footnotesize{}$15$} &  & {\footnotesize{}$\underset{\left(573\right)}{8.30}$} & {\footnotesize{}$\underset{\left(3808\right)}{8.34}$} & {\footnotesize{}$\underset{\left(6466\right)}{8.42}$} & {\footnotesize{}$\underset{\left(10222\right)}{8.45}$} &  & {\footnotesize{}$\underset{\left(1361\right)}{8.44}$} & {\footnotesize{}$\underset{\left(9213\right)}{8.46}$} & {\footnotesize{}$\underset{\left(12488\right)}{8.49}$} & {\footnotesize{}$\underset{\left(11531\right)}{8.51}$} &  & \tabularnewline
\midrule
{\small{}$110$} & {\footnotesize{}$0$} &  & {\footnotesize{}$\underset{\left(84\right)}{13.05}$} & {\footnotesize{}$\underset{\left(229\right)}{13.07}$} & {\footnotesize{}$\underset{\left(325\right)}{13.10}$} & {\footnotesize{}$\underset{\left(519\right)}{13.10}$} &  & {\footnotesize{}$\underset{\left(216\right)}{13.20}$} & {\footnotesize{}$\underset{\left(486\right)}{13.18}$} & {\footnotesize{}$\underset{\left(646\right)}{13.17}$} & {\footnotesize{}$\underset{\left(1048\right)}{13.17}$} &  & \multirow{5}{*}{{\scriptsize{}13.23}}\tabularnewline
 & {\footnotesize{}$1$} &  & {\footnotesize{}$\underset{\left(190\right)}{13.08}$} & {\footnotesize{}$\underset{\left(444\right)}{13.09}$} & {\footnotesize{}$\underset{\left(476\right)}{13.12}$} & {\footnotesize{}$\underset{\left(796\right)}{13.14}$} &  & {\footnotesize{}$\underset{\left(182\right)}{13.20}$} & {\footnotesize{}$\underset{\left(737\right)}{13.18}$} & {\footnotesize{}$\underset{\left(857\right)}{13.17}$} & {\footnotesize{}$\underset{\left(1770\right)}{13.17}$} &  & \tabularnewline
 & {\footnotesize{}$3$} &  & {\footnotesize{}$\underset{\left(180\right)}{13.06}$} & {\footnotesize{}$\underset{\left(728\right)}{13.08}$} & {\footnotesize{}$\underset{\left(1162\right)}{13.17}$} & {\footnotesize{}$\underset{\left(1111\right)}{13.10}$} &  & {\footnotesize{}$\underset{\left(454\right)}{13.24}$} & {\footnotesize{}$\underset{\left(1635\right)}{13.20}$} & {\footnotesize{}$\underset{\left(2035\right)}{13.21}$} & {\footnotesize{}$\underset{\left(2751\right)}{13.27}$} &  & \tabularnewline
 & {\footnotesize{}$7$} &  & {\footnotesize{}$\underset{\left(360\right)}{13.05}$} & {\footnotesize{}$\underset{\left(4208\right)}{13.16}$} & {\footnotesize{}$\underset{\left(2252\right)}{13.13}$} & {\footnotesize{}$\underset{\left(2111\right)}{13.13}$} &  & {\footnotesize{}$\underset{\left(772\right)}{13.20}$} & {\footnotesize{}$\underset{\left(4496\right)}{13.25}$} & {\footnotesize{}$\underset{\left(4532\right)}{13.24}$} & {\footnotesize{}$\underset{\left(4336\right)}{13.21}$} &  & \tabularnewline
 & {\footnotesize{}$15$} &  & {\footnotesize{}$\underset{\left(812\right)}{13.05}$} & {\footnotesize{}$\underset{\left(5221\right)}{13.09}$} & {\footnotesize{}$\underset{\left(6290\right)}{13.19}$} & {\footnotesize{}$\underset{\left(5257\right)}{13.12}$} &  & {\footnotesize{}$\underset{\left(2118\right)}{13.27}$} & {\footnotesize{}$\underset{\left(9895\right)}{13.21}$} & {\footnotesize{}$\underset{\left(13941\right)}{13.28}$} & {\footnotesize{}$\underset{\left(6948\right)}{13.22}$} &  & \tabularnewline
\midrule
{\small{}$120$} & {\footnotesize{}$0$} &  & {\footnotesize{}$\underset{\left(86\right)}{20.19}$} & {\footnotesize{}$\underset{\left(281\right)}{20.20}$} & {\footnotesize{}$\underset{\left(307\right)}{20.21}$} & {\footnotesize{}$\underset{\left(704\right)}{20.21}$} &  & {\footnotesize{}$\underset{\left(174\right)}{20.24}$} & {\footnotesize{}$\underset{\left(535\right)}{20.21}$} & {\footnotesize{}$\underset{\left(620\right)}{20.21}$} & {\footnotesize{}$\underset{\left(1087\right)}{20.21}$} &  & \multirow{5}{*}{{\scriptsize{}20.00}}\tabularnewline
 & {\footnotesize{}$1$} &  & {\footnotesize{}$\underset{\left(93\right)}{20.19}$} & {\footnotesize{}$\underset{\left(372\right)}{20.20}$} & {\footnotesize{}$\underset{\left(454\right)}{20.21}$} & {\footnotesize{}$\underset{\left(736\right)}{20.21}$} &  & {\footnotesize{}$\underset{\left(200\right)}{20.24}$} & {\footnotesize{}$\underset{\left(776\right)}{20.21}$} & {\footnotesize{}$\underset{\left(1025\right)}{20.22}$} & {\footnotesize{}$\underset{\left(1311\right)}{20.21}$} &  & \tabularnewline
 & {\footnotesize{}$3$} &  & {\footnotesize{}$\underset{\left(180\right)}{20.19}$} & {\footnotesize{}$\underset{\left(675\right)}{20.20}$} & {\footnotesize{}$\underset{\left(1002\right)}{20.20}$} & {\footnotesize{}$\underset{\left(1286\right)}{20.19}$} &  & {\footnotesize{}$\underset{\left(468\right)}{20.25}$} & {\footnotesize{}$\underset{\left(1825\right)}{20.22}$} & {\footnotesize{}$\underset{\left(1418\right)}{20.21}$} & {\footnotesize{}$\underset{\left(1971\right)}{20.21}$} &  & \tabularnewline
 & {\footnotesize{}$7$} &  & {\footnotesize{}$\underset{\left(323\right)}{20.19}$} & {\footnotesize{}$\underset{\left(2411\right)}{20.22}$} & {\footnotesize{}$\underset{\left(2307\right)}{20.21}$} & {\footnotesize{}$\underset{\left(3043\right)}{20.22}$} &  & {\footnotesize{}$\underset{\left(696\right)}{20.24}$} & {\footnotesize{}$\underset{\left(5008\right)}{20.20}$} & {\footnotesize{}$\underset{\left(2715\right)}{20.21}$} & {\footnotesize{}$\underset{\left(4496\right)}{20.19}$} &  & \tabularnewline
 & {\footnotesize{}$15$} &  & {\footnotesize{}$\underset{\left(1227\right)}{20.16}$} & {\footnotesize{}$\underset{\left(5759\right)}{20.20}$} & {\footnotesize{}$\underset{\left(3662\right)}{20.19}$} & {\footnotesize{}$\underset{\left(7173\right)}{20.21}$} &  & {\footnotesize{}$\underset{\left(1300\right)}{20.24}$} & {\footnotesize{}$\underset{\left(9185\right)}{20.22}$} & {\footnotesize{}$\underset{\left(5834\right)}{20.20}$} & {\footnotesize{}$\underset{\left(7580\right)}{20.21}$} &  & \tabularnewline
\midrule
{\small{}$130$} &  &  & \multicolumn{4}{c}{{\footnotesize{}Always $30.00$}} &  & \multicolumn{4}{c}{{\footnotesize{}Always $30.00$}} &  & {\scriptsize{}30.00}\tabularnewline
\midrule
{\small{}$140$} &  &  & \multicolumn{4}{c}{{\footnotesize{}Always $40.00$}} &  & \multicolumn{4}{c}{{\footnotesize{}Always $40.00$}} &  & {\scriptsize{}40.00}\tabularnewline
\bottomrule
\end{tabular}
\par\end{centering}
\caption{\label{tab:RB-EI}{\small{}Results for an American Put option in the
rough Bergomi model using the GPR-EI method. $N$ represents the number
of time steps, $P$ the number of the simulated paths and $J$ the
number of past values employed in the regression. Values in brackets
are the computational times (in seconds).}}
\end{table}

\FloatBarrier

\section{Conclusions}

In this paper we have presented two numerical methods to compute the
price of American options on a basket of underlyings following the
Black-Scholes dynamics. These two methods are based on the GPR-Monte
Carlo method and improve its results in terms of accuracy and computational
time. The GPR-Tree method can be applied for dimensions up to $d=20$
and it proves to be very efficient when $d\leq10$. The GPR-Exact
Integration method proves to be particularly flexible and stands out
for the small computational cost which allows one to obtain excellent
estimates in a very short time. The two methods also turns out to
be an effective tool to address non-Markovian problems such as the
pricing of American options in the rough Bergomi model. These two
methods are thus a step forward in overcoming the curse of dimensionality.

\bibliographystyle{abbrv}
\bibliography{bibliography}

\appendix

\section{\label{ApA0}Proof of Proposition \ref{prop:0}}

Let $n\in\left\{ 0,\dots,N-1\right\} $ and suppose the function $u\left(t_{n+1},\cdot\right)$
at time $t_{n+1}$ to be known at $Z$. Let us define the quantity
\begin{equation}
\hat{\mathbf{x}}^{p}=\exp\left(\mathbf{z}^{p}+\left(r-\frac{1}{2}\boldsymbol{\sigma}^{2}\right)t_{n}\right)
\end{equation}
for $p=1,\dots,P$. The function $u\left(t_{n},\cdot\right)$ at time
$t_{n}$ at $\mathbf{z}^{p}$ follows 
\begin{align}
u\left(t_{n},\mathbf{z}^{p}\right) & =v\left(t_{n},\hat{\mathbf{x}}^{p}\right).\\
 & =\max\left(\Psi\left(\hat{\mathbf{x}}^{p}\right),C\left(t_{n},\hat{\mathbf{x}}^{p}\right)\right),
\end{align}
where 
\begin{equation}
C\left(t_{n},\hat{\mathbf{x}}^{p}\right)=\mathbb{E}_{t_{n},\hat{\mathbf{x}}^{p}}\left[e^{-r\Delta t}v\left(t_{n+1},\mathbf{S}_{t_{n+1}}\right)\right]
\end{equation}
We can also write 
\begin{align}
C\left(t_{n},\hat{\mathbf{x}}^{p}\right) & =\mathbb{E}_{t_{n},\hat{\mathbf{x}}^{p}}\left[e^{-r\Delta t}u\left(t_{n+1},\log\left(\mathbf{S}_{t_{n+1}}\right)-\left(r-\frac{1}{2}\boldsymbol{\sigma}^{2}\right)t_{n+1}\right)\right]\\
 & =\mathbb{E}_{t_{n},\hat{\mathbf{x}}^{p}}\left[e^{-r\Delta t}u\left(t_{n+1},\mathbf{Z}_{t_{n+1}}\right)\right]\label{eq:A6a}
\end{align}
where $\mathbf{Z}_{t_{n+1}}$ is the random variable defined as 
\begin{equation}
\mathbf{Z}_{t_{n+1}}=\log\left(\mathbf{S}_{t_{n+1}}\right)-\left(r-\frac{1}{2}\boldsymbol{\sigma}^{2}\right)t_{n+1}.
\end{equation}
Let us define $\Pi=\left(\Pi_{i,j}\right)$ as the $d\times d$ covariance
matrix of the log-increments, that is $\Pi_{i,j}=\rho_{i,j}\sigma_{i}\sigma_{j}\Delta t$
. Moreover, let $\Lambda$ be a square root of $\Pi$ and $\mathbf{G}$
as a vector that follows a standard Gaussian law. Then, we observe
that $\mathbf{Z}_{t_{n+1}}$ has the following conditional law 
\begin{equation}
\mathbf{Z}_{t_{n+1}}\left|\mathbf{S}_{t_{n}}=\hat{\mathbf{x}}^{p}\right.\sim\mathcal{N}\left(\mathbf{z}^{p},\varPi\right).\label{eq:A6}
\end{equation}
In fact, simple Algebra leads to
\begin{equation}
\mathbf{Z}_{t_{n+1}}=\mathbf{z}^{p}+\Lambda\mathbf{G}.
\end{equation}
Moreover, relation (\ref{eq:A6}) can also be stated as
\begin{equation}
\mathbf{Z}_{t_{n+1}}\left|\left(\log\left(\mathbf{S}_{t_{n}}\right)-\left(r-\frac{1}{2}\boldsymbol{\sigma}^{2}\right)t_{n}=\mathbf{z}^{p}\right)\right.\sim\mathcal{N}\left(\mathbf{z}^{p},\varPi\right).
\end{equation}
Let $f_{\mathbf{z}^{p}}\left(\mathbf{z}\right)$ denote the density
function of $\mathbf{Z}_{t_{n+1}}$ given $\log\left(\mathbf{S}_{t_{n}}\right)-\left(r-\frac{1}{2}\boldsymbol{\sigma}^{2}\right)t_{n}=\mathbf{z}^{p}$
. Specifically,
\begin{equation}
f_{\mathbf{z}^{p}}\left(\mathbf{z}\right)=\frac{1}{\left(2\pi\right)^{\frac{d}{2}}\sqrt{\det\left(\Pi\right)}}\exp\left(-\frac{1}{2}\left(\mathbf{z}-\mathbf{z}^{p}\right)^{\top}\Pi^{-1}\left(\mathbf{\mathbf{z}}-\mathbf{z}^{p}\right)\right).
\end{equation}
Then, according to (\ref{eq:A6a}),we can write 
\begin{equation}
C\left(t_{n},\hat{\mathbf{x}}^{p}\right)=e^{-r\Delta t}\int_{\mathbb{R}^{d}}f_{\mathbf{z}^{p}}\left(\mathbf{z}\right)u\left(t_{n+1},\mathbf{z}\right)d\mathbf{z}.
\end{equation}

Now, let us consider GPR approximation of the function $u\left(t_{n+1},\cdot\right)$,
obtained by assuming $Z$  as the predictor set and by employing the
Squared Exponential Kernel $k_{SE}:\mathbb{R}^{d}\times\mathbb{R}^{d}\rightarrow\mathbb{R}$,
which is given by
\begin{equation}
k_{SE}\left(\mathbf{a},\mathbf{b}\right)=\sigma_{f}^{2}\exp\left(-\frac{\left(\mathbf{a}-\mathbf{b}\right)^{\top}I_{d}\left(\mathbf{a}-\mathbf{b}\right)}{2\sigma_{l}^{2}}\right),\ \mathbf{a},\mathbf{b}\in\mathbb{R}^{d}.\label{eq:A11}
\end{equation}
In particular, with reference to (\ref{eq:A11}), the additional parameters
$\sigma_{l}$ and $\sigma_{f}$ are called hyperparameters and are
obtained by means of a maximum likelihood estimation. So let 
\begin{equation}
u_{n+1}^{GPR}\left(\mathbf{z}\right)=\sum_{q=1}^{P}\mathbf{\omega}_{q}k_{SE}\left(\mathbf{z}^{q},\mathbf{z}\right),\label{eq:A12}
\end{equation}
be the GPR approximation of the function $u\left(t_{n+1},\mathbf{z}\right)$,
where $\boldsymbol{\omega}=\left(\omega_{1},\dots,\omega_{q},\dots\omega_{P}\right)^{\top}$
in (\ref{eq:A12}) is a vector of weights that can be computed by
solving a linear system (see Rasmussen and Williams \cite{williams2006gaussian}).
The GPR-EI approximation $C_{n}^{GPR-EI}$ of the continuation value
is then given by 
\begin{align}
C_{n}^{GPR-EI}\left(\hat{\mathbf{x}}^{p}\right) & =e^{-r\Delta t}\int_{\mathbb{R}^{d}}f_{\mathbf{z}^{p}}\left(\mathbf{z}\right)u_{n+1}^{GPR}\left(\mathbf{z}\right)d\mathbf{z}\\
 & =e^{-r\Delta t}\sum_{q=1}^{P}\omega_{q}\int_{\mathbb{R}^{d}}f_{\mathbf{z}^{p}}\left(\mathbf{z}\right)k_{SE}\left(\mathbf{z}^{q},\mathbf{z}\right)d\mathbf{z}.\label{eq:A14}
\end{align}
To compute each integral in (\ref{eq:A14}), we observe that 
\begin{multline}
\int_{\mathbb{R}^{d}}f_{\mathbf{z}^{p}}\left(\mathbf{z}\right)k_{SE}\left(\mathbf{z}^{q},\mathbf{z}\right)d\mathbf{z}=\\
=\left(2\pi\right)^{\frac{d}{2}}\sigma_{f}^{2}\sigma_{l}^{d}\int_{\mathbb{R}^{d}}\frac{1}{\left(2\pi\right)^{\frac{d}{2}}\sqrt{\det\left(\Pi\right)}}e^{-\frac{1}{2}\left(\mathbf{z}-\mathbf{z}^{p}\right)^{\top}\Pi^{-1}\left(\mathbf{\mathbf{z}}-\mathbf{z}^{p}\right)}\frac{1}{\left(2\pi\right)^{\frac{d}{2}}\sqrt{\sigma_{l}^{2d}}}e^{-\frac{1}{2}\left(\mathbf{z}-\mathbf{z}^{q}\right)^{\top}\left(\sigma_{l}^{2}I_{d}\right)^{-1}\left(\mathbf{z}-\mathbf{z}^{q}\right)}d\mathbf{z}\\
=\left(2\pi\right)^{\frac{d}{2}}\sigma_{f}^{2}\sigma_{l}^{d}\int_{\mathbb{R}^{d}}\frac{1}{\left(2\pi\right)^{\frac{d}{2}}\sqrt{\det\left(\Pi\right)}}e^{-\frac{1}{2}\left(\mathbf{z}-\mathbf{z}^{p}\right)^{\top}\Pi^{-1}\left(\mathbf{\mathbf{z}}-\mathbf{z}^{p}\right)}\frac{1}{\left(2\pi\right)^{\frac{d}{2}}\sqrt{\sigma_{l}^{2d}}}e^{-\frac{1}{2}\left(\left(\mathbf{0}-z\right)-\left(-\mathbf{z}^{q}\right)\right)^{\top}\left(\sigma_{l}^{2}I_{d}\right)^{-1}\left(\left(\mathbf{0}-\mathbf{z}\right)-\left(-\mathbf{z}^{q}\right)\right)}d\mathbf{z}\\
=\left(2\pi\right)^{\frac{d}{2}}\sigma_{f}^{2}\sigma_{l}^{d}f_{\mathbf{z}^{p}}\ast g_{\mathbf{-z}^{q}}\left(0\right)
\end{multline}
where $\ast$ is the convolution product and $g_{\mathbf{-z}^{q}}$
is the density function of a Gaussian random vector which has law
given by $\mathcal{N}\left(-\mathbf{z}^{q},\sigma_{l}^{2}I_{d}\right)$.
Moreover, the convolution product of the densities of two independent
random variables is equal to the density of their sum (see Hogg et
al. \cite{hogg2005introduction}) and we can obtain the following
relation which allows one to exactly compute the integrals in (\ref{eq:A14}):
\begin{equation}
f_{\mathbf{x}^{p}}\ast g_{\mathbf{-x}^{q}}\left(0\right)=\frac{1}{\left(2\pi\right)^{\frac{d}{2}}\sqrt{\det\left(\Pi+\sigma_{l}^{2}I_{d}\right)}}e^{-\frac{1}{2}\left(\mathbf{z}^{q}-\mathbf{z}^{p}\right)^{\top}\left(\Pi+\sigma_{l}^{2}I_{d}\right)^{-1}\left(\mathbf{z}^{q}-\mathbf{z}^{p}\right)}.
\end{equation}
Therefore, the GPR-EI approximation $C_{n}^{GPR-EI}$ at $\hat{\mathbf{x}}^{p}$
reads
\begin{equation}
C_{n}^{GPR-EI}\left(\hat{\mathbf{x}}^{p}\right)=e^{-r\Delta t}\sum_{q=1}^{P}\omega_{q}\sigma_{f}^{2}\sigma_{l}^{d}\frac{e^{-\frac{1}{2}\left(\mathbf{z}^{q}-\mathbf{z}^{p}\right)^{\top}\left(\Pi+\sigma_{l}^{2}I_{d}\right)^{-1}\left(\mathbf{z}^{q}-\mathbf{z}^{p}\right)}}{\sqrt{\det\left(\Pi+\sigma_{l}^{2}I_{d}\right)}},
\end{equation}
and the GPR-EI approximation $u_{n}^{GPR-EI}$ of the option value
$u\left(t_{n},\cdot\right)$ at time $t_{n}$ and at $\mathbf{z}^{p}$
is given by

\begin{equation}
u_{n}^{GPR-EI}\left(\mathbf{z}^{p}\right)=\max\left(\Psi\left(\hat{\mathbf{x}}^{p}\right),e^{-r\Delta t}\sum_{q=1}^{P}\omega_{q}\sigma_{f}^{2}\sigma_{l}^{d}\frac{e^{-\frac{1}{2}\left(\mathbf{z}^{q}-\mathbf{z}^{p}\right)^{\top}\left(\Pi+\sigma_{l}^{2}I_{d}\right)^{-1}\left(\mathbf{z}^{q}-\mathbf{z}^{p}\right)}}{\sqrt{\det\left(\Pi+\sigma_{l}^{2}I_{d}\right)}}\right).\label{eq:v_GPR-EI}
\end{equation}

\section{\label{ApA}Covariance of the vector $R$ in (\ref{eq:vector_R})}

Let us report the formulas for the covariance of the components of
the vector $R$ in (\ref{eq:vector_R}). For all $n=1,\dots,N$, and
$m=1,\dots,n-1$, the following relations hold:
\begin{equation}
Cov\left(\Delta W_{n}^{1},\Delta W_{n}^{1}\right)=\Delta t,
\end{equation}

\begin{equation}
Cov\left(\Delta W_{n}^{1},\widetilde{W}_{t_{n}}^{H}\right)=\frac{2\rho\sqrt{2H}}{2H+1}\left(\Delta t\right)^{H+\frac{1}{2}},
\end{equation}
\begin{equation}
Cov\left(\widetilde{W}_{t_{n}}^{H},\widetilde{W}_{t_{n}}^{H}\right)=\left(t_{n}\right)^{2H}
\end{equation}

\begin{equation}
Cov\left(\Delta W_{m}^{1},\Delta W_{n}^{1}\right)=0,
\end{equation}

\begin{equation}
Cov\left(\Delta W_{n}^{1},\widetilde{W}_{t_{m}}^{H}\right)=0,
\end{equation}

\begin{equation}
Cov\left(\Delta W_{m}^{1},\widetilde{W}_{t_{n}}^{H}\right)=\frac{2\rho\sqrt{2H}}{2H+1}\left(\left(t_{n}-t_{m-1}\right)^{H+\frac{1}{2}}-\left(t_{n}-t_{m}\right)^{H+\frac{1}{2}}\right),
\end{equation}

\begin{equation}
Cov\left(\widetilde{W}_{t_{m}}^{H},\widetilde{W}_{t_{n}}^{H}\right)=2H\left(t_{m}\right)^{2H}\cdot\int_{0}^{1}\frac{ds}{\left(1-s\right)^{\frac{1}{2}-H}\left(\frac{t_{m}}{t_{n}}-s\right)^{\frac{1}{2}-H}}.
\end{equation}

\section{\label{ApA2}Proof of Proposition \ref{lem:L1}}

Let us denote the random vector $\left(S_{t_{i}},V_{t_{i}},S_{t_{i+1}},V_{t_{i+1}},\dots,S_{t_{j}},V_{t_{j}}\right)^{\top}$
for $i,j\in\left\{ 0,\dots,N\right\} $ and $i<j$ with $\mathbf{SV}_{i:j}$
. We observe that the option value $v\left(t_{N},\cdot\right)$ at
time $t_{N}$ is given by the payoff function $\Psi$, which only
depends by the final value of the underlying. The option value $v\left(t_{N-1},\cdot\right)$
at time $t_{N-1}$ about the $p$-th path is given by 
\begin{equation}
v\left(t_{N-1},\mathbf{SV}_{1:\left(N-1\right)}^{p}\right)=\max\left(\Psi\left(S_{t_{N-1}}^{p}\right),e^{-r\Delta t}C\left(t_{N-1},\mathbf{SV}_{1:\left(N-1\right)}^{p}\right)\right)\label{eq:65}
\end{equation}
where $C$ stands for the continuation value and it is equal to 
\begin{equation}
C\left(t_{N-1},\mathbf{SV}_{1:\left(N-1\right)}^{p}\right)=E\left[e^{-r\Delta t}\Psi\left(S_{t_{N}}\right)\left|\left(\mathbf{SV}_{1:\left(N-1\right)}=\mathbf{SV}_{1:\left(N-1\right)}^{p}\right)\right.\right].\label{eq:CVR}
\end{equation}
We approximate the continuation value in (\ref{eq:CVR}) by means
of the GPR approximation of $\Psi$. In particular, let $\Psi^{GPR}\left(z\right)$
be the approximation of the function $z\mapsto\Psi\left(\exp\left(z\right)\right)$
by using the GPR method employing the Squared Exponential Kernel and
considering the log-underlying values at maturity as predictors. Specifically,
the predictor set is
\begin{equation}
Z=\left\{ z^{p}=\log\left(S_{t_{N}}^{p}\right),p=1,\dots,P\right\} \subset\mathbb{R}^ {}
\end{equation}
and the response $\mathbf{y}\in\mathbb{R}^{P}$ is given by 
\begin{equation}
y^{p}=\Psi\left(S_{t_{N}}^{p}\right).
\end{equation}
 In particular, we can write 
\begin{align}
\Psi^{GPR}\left(z\right) & =\sum_{q=1}^{P}k_{SE}\left(\log\left(S_{t_{N}}^{q}\right),z\right)\mathbf{\omega}_{q}=\sigma_{f}^{2}\sum_{q=1}^{P}\exp\left(-\frac{\left(\log\left(S_{t_{N}}^{q}\right)-z\right)^{2}}{2\sigma_{l}^{2}}\right)\mathbf{\omega}_{q}
\end{align}
where $k_{SE}$ is the Squared Exponential kernel, $\sigma_{l}$ is
the characteristic length scale, $\sigma_{f}$ is the signal standard
deviation and $\omega_{1},\dots,\omega_{P}$ are weights. 

So we approximate the continuation value $C\left(t_{N-1},\mathbf{SV}_{1:\left(N-1\right)}^{p}\right)$
with the expression:
\begin{equation}
E\left[e^{-r\Delta t}\Psi^{GPR}\left(\ln\left(S_{t_{N}}\right)\right)\left|\left(\mathbf{SV}_{1:\left(N-1\right)}=\mathbf{SV}_{1:\left(N-1\right)}^{p}\right)\right.\right].
\end{equation}
We observe that the law of $\log\left(S_{t_{N}}\right)$ given $S_{t_{1}}^{p},V_{t_{1}}^{p},\dots,S_{t_{N-1}}^{p},V_{t_{N-1}}^{p}$
is normal
\begin{equation}
\log\left(S_{t_{N}}\right)\left|\left(\mathbf{SV}_{1:\left(N-1\right)}=\mathbf{SV}_{1:\left(N-1\right)}^{p}\right)\right.\sim\mathcal{N}\left(\mu_{N,p},\sigma_{N,p}^{2}\right),
\end{equation}
where 
\begin{equation}
\mu_{N,p}=\log\left(S_{t_{N-1}}^{p}\right)+\left(r-\frac{1}{2}V_{t_{N-1}}^{p}\right)\Delta t
\end{equation}
and 
\begin{equation}
\sigma_{N,p}^{2}=V_{t_{N-1}}^{p}\Delta t.
\end{equation}
Therefore, the GPR-EI approximation for the continuation value at
time $t_{N-1}$ is as follows:
\begin{multline}
C_{N-1}^{GPR-EI}\left(\mathbf{SV}_{1:\left(N-1\right)}^{p}\right)=e^{-r\Delta t}\int_{\mathbb{R}}\frac{\exp\left(-\frac{\left(z-\mu_{N,p}\right)^{2}}{2\sigma_{N,p}^{2}}\right)}{\sqrt{2\pi\sigma_{N,p}^{2}}}\Psi^{GPR}\left(z\right)dz\\
=e^{-r\Delta t}\sigma_{f}^{2}\sqrt{2\pi\sigma_{l}^{2}}\sum_{q=1}^{P}\int_{\mathbb{R}}\frac{\exp\left(-\frac{\left(z-\mu_{N,p}\right)^{2}}{2\sigma_{N,p}^{2}}\right)}{\sqrt{2\pi\sigma_{N,p}^{2}}}\frac{\exp\left(-\frac{\left(\log\left(S_{t_{N}}^{q}\right)-z\right)^{2}}{2\sigma_{l}^{2}}\right)}{\sqrt{2\pi\sigma_{l}^{2}}}\mathbf{\omega}_{q}dz.
\end{multline}
Taking advantage of the properties of the convolution between density
functions, we obtain 
\begin{equation}
C_{N-1}^{GPR-EI}\left(\mathbf{SV}_{1:\left(N-1\right)}^{p}\right)=e^{-r\Delta t}\sum_{q=1}^{P}\frac{\mathbf{\omega}_{q}\sigma_{f}^{2}\sigma_{l}}{\sqrt{\sigma_{N,p}^{2}+\sigma_{l}^{2}}}\exp\left(-\frac{\left(\log\left(S_{t_{N}}^{q}\right)-\mu_{N,p}\right)^{2}}{2\sigma_{N,p}^{2}+2\sigma_{l}^{2}}\right).\label{eq:CVNm1}
\end{equation}

\section{\label{ApA3}Proof of Proposition \ref{lem:L2}}

In order to proceed backward, from $t_{N-2}$ up to $t_{1}$ we consider
an integer positive value $J$ and train the GPR method considering
the last $J+1$ observed values of the couple $\left(\log\left(S_{t_{n}}^{p}\right),\log\left(V_{t_{n}}^{p}\right)\right)$
as predictors, and the option price as response. Specifically, the
predictor set is
\begin{equation}
Z=\left\{ \mathbf{z}^{p}=\log\left(\mathbf{SV}_{\max\left\{ 1,N-1-J\right\} :\left(N-1\right)}^{p}\right),p=1,\dots,P\right\} \subset\mathbb{R}^{d_{N-1}}
\end{equation}
where $d_{N-1}=2\min\left\{ N-1,J+1\right\} $ and the response $\mathbf{y}\in\mathbb{R}^{P}$
is given by 
\begin{equation}
y^{p}=v\left(t_{N-1},\mathbf{SV}_{1:\left(N-1\right)}^{p}\right).
\end{equation}
We term $u_{N-1}^{GPR}$ the obtained function. In particular, $u_{N-1}^{GPR}:\mathbb{R}^{d_{N-1}}\rightarrow\mathbb{R}$
and
\begin{equation}
u_{N-1}^{GPR}\left(\log\left(\mathbf{SV}_{\max\left\{ 1,N-1-J\right\} :\left(N-1\right)}^{p}\right)\right)
\end{equation}
 approximates $v\left(t_{N-1},\mathbf{SV}_{1:\left(N-1\right)}^{p}\right)$. 

Since the predictors have different nature (log-prices and log-volatilities
at different times), we use the Automatic Relevance Determination
(ARD) Squared Exponential Kernel $k_{ASE}$ to perform the GPR regression.
In particular, if $d$ is the dimension of the space containing the
predictors, it holds
\begin{equation}
k_{ASE}\left(\mathbf{a},\mathbf{b}\right)=\sigma_{f}^{2}\exp\left(-\sum_{i=1}^{d}\frac{\left(a_{i}-b_{i}\right)^{2}}{2\sigma_{i}^{2}}\right),\ \mathbf{a},\mathbf{b}\in\mathbb{R}^{d},
\end{equation}
As opposed to the Squared Exponential kernel, the ARD Squared Exponential
kernel considers a different length scale $\sigma_{i}$ for each predictor
that allows the regression to better learn the impact of each predictor
on the response.

We present now how to perform the backward induction. So, let us consider
$n\in\left\{ 0,\dots,N-2\right\} $ and suppose the GPR approximation
$u_{n+1}^{GPR}:\mathbb{R}^{d_{n+1}}\rightarrow\mathbb{R}$ to be known.
In particular, $d_{n+1}=2\min\left\{ n+1,J+1\right\} $ and for each
$\mathbf{z}=\left(z_{1},\dots,z_{d_{n+1}}\right)\in\mathbb{R}^{d_{n+1}},$
it holds
\begin{equation}
u_{n+1}^{GPR}\left(\mathbf{z}\right)=\sigma_{f}^{2}\sum_{q=1}^{P}\mathbf{\omega}_{q}\exp\left(-\sum_{i=1}^{d_{n}}\frac{\left(z_{i}^{q}-z_{i}\right)^{2}}{2\sigma_{i}^{2}}\right),
\end{equation}
where $z_{i}^{q}=\log\left(S_{n+1-\left(i-1\right)/2}^{q}\right)$
if $i$ is even and $z_{i}^{q}=\log\left(V_{n+1-i/2}^{q}\right)$
if $i$ is odd, for $i=1,\dots,d_{n+1}$. This means that $z_{i}^{q}$
is the observed log-price at time $t_{n+1-\left(i-1\right)/2}$ of
the $q$-th path if $i$ is even, and it is the observed log-volatility
at time $t_{n+1-\left(i-1\right)/2}$ of the $q$-th path if $i$
is odd. 

We explain now how to compute the GPR approximation $v_{n}^{GPR-EI}:\mathbb{R}^{d_{n}}\rightarrow\mathbb{R}$
of the price function at time $t_{n}$. First of all, we observe that
the vector $\left(\log\left(S_{t_{n+1}}^{p}\right),\log\left(V_{t_{n+1}}^{p}\right)\right)^{\top}$
is not $\hat{\mathcal{F}}_{t_{n}}$-measurable whereas $\log\left(\mathbf{SV}_{\max\left\{ 1,n+1-J\right\} :n}^{p}\right)$
is $\hat{\mathcal{F}}_{t_{n}}$-measurable. The law of $\left(\log\left(S_{t_{n+1}}\right),\log\left(V_{t_{n+1}}\right)\right)^{\top}$
given $S_{t_{n}}^{p},V_{t_{n}}^{p},\dots,S_{t_{1}}^{p},V_{t_{1}}^{p}$
is normal:
\begin{equation}
\left(\log\left(S_{t_{n+1}}\right),\log\left(V_{t_{n+1}}\right)\right)^{\top}\left|\left(\mathbf{SV}_{1:n}=\mathbf{SV}_{1:n}^{p}\right)\right.\sim\mathcal{N}\left(\mu_{n+1,p},\Sigma_{n+1,p}\right),
\end{equation}
 In particular
\begin{equation}
\mu_{n+1,p}=\left(\log\left(S_{t_{n}}^{p}\right)+\left(r-\frac{1}{2}V_{t_{n}}^{p}\right)\Delta t,\log\left(\xi_{0}\right)+\eta\Lambda_{2n+2}\underline{\mathbf{G}}^{p}-\frac{1}{2}\eta^{2}t_{n+1}^{2H}\right)^{\top},
\end{equation}
where $\Lambda_{2n+2}$ is the $2n+2$-th row of the matrix $\Lambda$
and $\underline{\mathbf{G}}^{p}=\left(G_{1}^{p},\dots,G_{2n}^{p},0\dots,0\right)^{\top}$.
Moreover, the covariance matrix is given by
\begin{equation}
\Sigma_{n+1,p}=\left(\begin{array}{cc}
\Delta tV_{t_{n}}^{p} & \eta\sqrt{\Delta tV_{t_{n}}^{p}}\Lambda_{2n+2,2n+1}\\
\eta\sqrt{\Delta tV_{t_{n}}^{p}}\Lambda_{2n+2,2n+1} & \eta^{2}\left(\Lambda_{2n+2,2n+2}^{2}+\Lambda_{2n+2,2n+1}^{2}\right)
\end{array}\right),
\end{equation}
where $\Lambda_{i,j}$ stands for the element of $\Lambda$ in position
$i,j$. Using a similar reasoning as done for the continuation value
at time $t_{N-1}$, one can obtain the following GPR-EI approximation
for the continuation value at time $t_{n-1}$:

\begin{equation}
C_{n}^{GPR-EI}\left(\mathbf{SV}_{\max\left\{ 1,n-J\right\} :n}^{p}\right)=e^{-r\Delta t}\sigma_{f}^{2}\sigma_{d_{n+1}-1}\sigma_{d_{n+1}}\sum_{q=1}^{P}\mathbf{\omega}_{q}h_{q}^{p}f_{q}^{p},\label{eq:622}
\end{equation}
where $h_{q}^{p}$ and $f_{q}^{p}$ are two factors given by
\begin{equation}
h_{q}^{p}=\exp\left(-\sum_{i=1}^{d_{n+1}-2}\frac{\left(z_{i}^{p}-z_{i}^{q}\right)^{2}}{2\sigma_{i}^{2}}\right)
\end{equation}
and

\begin{equation}
f_{q}^{p}=\frac{\exp\left(-\frac{1}{2}\left(\left(\begin{array}{c}
z_{d_{n+1}-1}^{q}\\
z_{d_{n+1}}^{q}
\end{array}\right)-\mu_{n+1,p}\right)^{\top}\left(\Sigma_{n+1,p}+\left(\begin{array}{cc}
\sigma_{d_{n+1}-1}^{2} & 0\\
0 & \sigma_{d_{n+1}}^{2}
\end{array}\right)\right)^{-1}\left(\left(\begin{array}{c}
z_{d_{n+1}-1}^{q}\\
z_{d_{n+1}}^{q}
\end{array}\right)-\mu_{n+1,p}\right)\right)}{\sqrt{\text{\ensuremath{\det}}\left(\Sigma_{n+1,p}+\left(\begin{array}{cc}
\sigma_{d_{n+1}-1}^{2} & 0\\
0 & \sigma_{d_{n+1}}^{2}
\end{array}\right)\right)}}.
\end{equation}
In particular, $h_{q}^{p}$ measures the impact of the past observed
values on the price, whereas $f_{q}^{p}$ integrates the changes due
to the diffusion of the underlying and its volatility.

Therefore, we obtain 
\[
v_{n}^{GPR-EI}\left(\mathbf{SV}_{\max\left\{ 1,n-J\right\} :n}^{p}\right)=\max\left(\Psi\left(S_{t_{n}}^{p}\right),e^{-r\Delta t}\sigma_{f}^{2}\sigma_{d_{n+1}-1}\sigma_{d_{n+1}}\sum_{q=1}^{P}\mathbf{\omega}_{q}h_{q}^{p}f_{q}^{p}\right).
\]
Finally, we observe that, in order to compute $u_{n}^{GPR}$, we train
the GPR method considering the predictor set given by
\begin{equation}
Z=\left\{ \mathbf{z}^{p}=\log\left(\mathbf{SV}_{\max\left\{ 1,n-J\right\} :n}^{p}\right),p=1,\dots,P\right\} \subset\mathbb{R}^{d_{n}}
\end{equation}
and the response $\mathbf{y}\in\mathbb{R}^{P}$ is given by 
\begin{equation}
y^{p}=v_{n}^{GPR-EI}\left(\mathbf{SV}_{\max\left\{ 1,n-J\right\} :n}^{p}\right).
\end{equation}

By induction we can compute the option price value for $n=N-2,\dots,0$
. 

To conclude, we observe that the continuation value at time $t=0$
can be computed by using (\ref{eq:622}) and considering $h_{q}^{p}=1$
for $q=1,\dots,P$ since in this case, there are no past values to
consider.
\end{document}